\documentstyle[aps,twocolumn,eqsecnum,psfig]{revtex}
\begin{document}
\draft
\wideabs{
\title{General-relativistic coupling between orbital motion and
internal degrees of freedom for inspiraling binary neutron stars.}

\author{\'Eanna \'E.\ Flanagan}
\address{Cornell University, Newman Laboratory, Ithaca, NY
14853-5001.}
\date{draft of \today}
\maketitle

%\narrowtext
%\twocolumn

\begin{abstract}
We analyze the coupling between the internal degrees of freedom of
neutron stars in a close binary, and the stars' orbital motion.  Our analysis
is based on the method of matched asymptotic expansions and is valid
to all orders in the strength of internal gravity in each star, but
is perturbative in the ``tidal expansion parameter'' (stellar
radius)/(orbital separation).  At first order in the tidal expansion
parameter, we show that the internal structure of each star is
unaffected by its companion, in agreement with post-1-Newtonian
results of Wiseman (gr-qc/9704018).  We also show
that relativistic interactions that scale as higher powers of the
tidal expansion parameter produce qualitatively similar effects to
their Newtonian counterparts: there are corrections to
the Newtonian tidal distortion of each star, both of which occur at
third order in the tidal expansion parameter, and there are
corrections to the Newtonian decrease in central
density of each star (Newtonian ``tidal stabilization''), both of
which are sixth order in the tidal expansion parameter.
There are additional interactions with no Newtonian analogs, but these
do not change the central density of each star up to sixth order in
the tidal expansion parameter.
These results, in combination with previous analyses of Newtonian tidal
interactions, indicate that (i) there are no large
general-relativistic crushing forces that could cause the stars to
collapse to black holes prior to
the dynamical orbital instability, and (ii) the conventional wisdom
with respect to coalescing binary neutron stars as sources of
gravitational-wave bursts is correct: namely, the finite-stellar-size
corrections to the gravitational waveform will be unimportant for the
purpose of detecting the coalescences.
\end{abstract}

}

\def\beq{\begin{equation}}
\def\endeq{\end{equation}}

%\newpage
%\twocolumn

\section{INTRODUCTION AND SUMMARY}
\label{sec:intro}

Recent numerical simulations by Wilson, Mathews and Marronetti of the
late stages of inspiral of neutron star binaries have predicted the
following surprising result:  The individual neutron stars are
apparently subject to a crushing force of general-relativistic origin
which can cause the stars to collapse and form black holes, before they
reach the dynamical orbital instability that marks the end of the
inspiral \cite{wmm,ReturnOfTheJedi}.  These numerical
simulations were fully
relativistic, but assumed a conformally flat spatial metric,
and also employed an approximation scheme in which the gravitational
field was constrained to be time-symmetric at each time-step in the
computation.  These approximations and assumptions give correct
results for spherically symmetric systems and also to the first
post-Newtonian approximation \cite{Reith}; beyond this, however, their
domain of validity is not well understood.

The Wilson-Mathews-Marronetti prediction is in disagreement with
other, independent, fully relativistic, numerical simulations
which employ similar approximations \cite{saul,stu1}, with post-1-Newtonian
numerical simulations \cite{Shibata},
and with Post-Newtonian \cite{lai,Wiseman,Lombardi} and
perturbation \cite{BradyHughes} calculations, which we discuss further below.
Therefore it seems likely that star crushing does not occur in reality,
although the issue is still somewhat controversial.

The star-crushing scenario, if correct, would have profound
implications for the efforts to detect gravitational waves
produced by neutron star inspirals with ground based interferometers such
as LIGO and VIRGO \cite{ligo}.  A crushing force which is strong enough to
cause an instability to radial collapse of a neutron star would
constitute a strong coupling between the orbital motion and the
internal modes of each star \cite{note88}, and would transfer a
substantial amount
of energy from the orbital motion to each
star.  Specifically, let $L_{\rm instability}$ be the value of the orbital
separation $L$ at the onset of instability, let $R$ be the initial radius of
the neutron star, and let $\Delta R$ be the amount by which the star's
radius is
decreased before the instability occurs.  Then the ratio of the
energy $\Delta E_{\rm star}$ absorbed by the neutron star \cite{notea}
to the energy $\Delta E_{\rm gw}$ radiated in gravitational waves
between $L = L_{\rm instability}/2$ and $L = L_{\rm instability}$ would be
approximately
\begin{equation}
{ \Delta E_{\rm star} \over \Delta E_{\rm gw}} \sim \left( { L_{\rm
instability} \over R} \right) \, \left( {\Delta R  \over R} \right).
\label{energyratio}
\end{equation}
The Wilson-Mathews-Marronetti simulations predict that an instability
occurs at $L_{\rm instability} \sim 5 R$, and that $\Delta R / R \sim
1/20$ \cite{note10}.
Therefore the ratio (\ref{energyratio}) is predicted to be of order
unity, and the crushing effect gives rise to an order unity
perturbation to the inspiral rate of the orbit and to the phase
evolution of the emitted
gravitational waves \cite{note3a}.  Since high accuracy theoretical
templates are
required in order to extract the signal from detector noise, the
star-crushing scenario would imply that currently envisaged search
templates \cite{searchtemplates} (which are calculated neglecting all
orbital-motion---internal-mode couplings) would need to be completely
revised.  Therefore, it is important to find out whether or not the star
crushing effect occurs.

In the Newtonian approximation, the coupling between the stars' orbital
motions and their internal motions has been analyzed in detail
\cite{BildstenCutler,Kochanek,LaiWiseman,Dong1,LRS,Kokkotas}.  The
Newtonian coupling
is very weak, too weak to affect the gravitational wave
signal except in the last few orbits before coalescence
\cite{BildstenCutler,Kochanek,LaiWiseman,Dong1,LRS,Kokkotas}.
Moreover, the
Newtonian interaction energy is $2 - 3$ orders of magnitude smaller
at $L \sim 5 R$ than the amount (\ref{energyratio}) which would be
required to crush
neutron stars when the necessary $\Delta R / R$ is of order several
percent \cite{lai,note10}.  Although post-Newtonian
couplings have not yet been analyzed in detail \cite{note11}, the
expectation has generally been that post-Newtonian or relativistic
couplings would simply be of the form (Newtonian coupling) $\times$
 $\ O(M/R)$ or $O(M/L)$, where $M$ is the neutron star mass.  That is,
only small fractional corrections to existing Newtonian couplings are
expected.
However, the Wilson-Mathews-Marronetti prediction suggested instead that
relativistic couplings could dominate over the Newtonian ones, and
highlighted the need to understand the details of these
couplings.  The purpose of this paper is to explore and
elucidate the relativistic, post-Newtonian couplings.

In the simulations of Wilson, Matthews and Marronetti \cite{wmm}, the central
density $\rho_c$ of each star was seen to increase during the
inspiral, and the instability to radial collapse occurred once the
fractional increase in central density reached a few percent.  In this
paper, we shall not consider the issue of stability to radial collapse, but
instead, following Refs.~\cite{Wiseman,BradyHughes}, we shall focus
attention on how the central density changes during the inspiral.
Focusing on the central density allows us to investigate the existence
of radial crushing forces (which presumably cause the instability to
collapse in the simulations).

\subsection{Relativistic coupling between orbital motion and internal
degrees of freedom in neutron star binaries}
\label{intro:coupling}

For a neutron star in a binary system, let $L$, $R$ and $M$ be the
orbital separation, stellar radius and stellar mass, respectively.
Then, there are two natural, independent, dimensionless parameters that
characterize the system: the strength of internal gravity in each star
\begin{equation}
\epsilon \equiv {M \over R},
\label{epsilondef}
\end{equation}
which is of order $\sim 0.2$, and the tidal expansion parameter
\begin{equation}
\alpha \equiv { R \over L}
\label{alphadef}
\end{equation}
which gradually increases during the inspiral.
In terms of these parameters, the post-Newtonian
expansion parameter for the orbital motion is $\epsilon_{\rm orbit} =
M/L = \epsilon \alpha$.  The standard post-Newtonian approximation scheme
consists of expanding in $\epsilon$ and $\epsilon_{\rm orbit}$,
treating these quantities as formally of the same order.
In the context of a neutron star binary, the usual terminology for
describing the size of gravitational effects
(Newtonian, Post-1-Newtonian, Post-2-Newtonian etc.) is somewhat
ambiguous, since an effect of post-$n$-Newtonian order could scale as
$\epsilon^a (\alpha \epsilon)^b$ for any $a$, $b$ with $a + b = n$.
In addition, some authors would classify an interaction
which scales like $\epsilon^s \alpha^t$ as being of effective
post-Newtonian order $t$, irrespective of the value of $s$, since the
index $t$ controls the perturbation to the phase evolution of the
emitted gravitational waves.  Therefore, in this paper we will not use the
post-Newtonian terminology.

Let us start by reviewing the situation in Newtonian gravity, at order
$\epsilon^0$ (see Table \ref{table1} below).  As is well known, each star
distorts the other at  $O(\alpha^3)$ via the quadrupole tidal interaction,
and the second order response of each star to the quadrupole tidal
field of its companion gives rise to a decrease in its central density
$\rho_c$ at $O(\alpha^6)$ \cite{lai}.  Each star expands slightly and
thus becomes more stable against radial collapse \cite{lai}.

Consider now the situation at higher orders in $\epsilon$.
If there were a relativistic crushing force that changed the central
density of each star, then this change in density $\delta \rho_c$
would scale in some
way with the dimensionless parameters $\alpha$ and $\epsilon$:
\begin{equation}
{\delta \rho_c \over \rho_c} \propto  \alpha^t \, \epsilon^s,
\label{guess1}
\end{equation}
for some integers or half-integers $s, t > 0$.  Such an effect would
dominate over the Newtonian effect at large orbital separations if $t
< 6$.  Note that naive arguments (which turn out to be incorrect) do
suggest scalings of the form (\ref{guess1}) with low powers of $\alpha$
\cite{explain}.  In the numerical
simulations it was seen that $\delta \rho_c / \rho_c \propto \alpha^2$
\cite{ReturnOfTheJedi}, while the scaling with $\epsilon$ was not clear.
A post-1-Newtonian calculation by Wiseman \cite{Wiseman} has shown that
there is no change in $\rho_c$ at the order $O(\alpha \, \epsilon)$, but
did not rule out the possibilities $\delta \rho_c /\rho_c \propto
\epsilon \, \alpha^t$ with $1 < t < 6$, or $\delta \rho_c/\rho_c \propto
\epsilon^s \, \alpha$ with $s >1$.

A different approximation scheme was used by Brady and Hughes to
investigate relativistic interactions \cite{BradyHughes}.
Suppose that the two stars are labeled A and B, and that their masses
and radii are $M_A$, $M_B$, $R_A$ and $R_B$.  Brady and
Hughes showed using perturbation theory that there is no
change in the central density of star A to linear order in
$M_B$.  However, this result would not rule out a change in central
density scaling as, for example,
\begin{equation}
{\delta \rho_c \over \rho_c} \ \propto \ {M_A M_B^2 \over R_A L^2}.
\label{counter-example}
\end{equation}
If we specialize to the equal mass case $M_A = M_B$ and $R_A = R_B$,
the scaling (\ref{counter-example}) reduces to $\delta \rho_c / \rho_c
\ \propto \ \epsilon \alpha^2$.  Thus, the Brady-Hughes analysis does
not necessarily rule out the type of behavior seen in the numerical
simulations.

In this paper we analyze the change in central density to all orders
in $\epsilon$, but perturbatively in $\alpha$.  We show that in a
binary system, the
excitation of the internal degrees of freedom of a fully relativistic
spherical star is qualitatively the same as that of a Newtonian
star.  At order $O(\alpha^3)$, the star
responds linearly to the external gravitational field, and is distorted; but
there is no excitation of the stars spherically symmetric,
radial modes.  At order $O(\alpha^6)$, the second-order response of
the star to the external field generates an excitation of the
stars spherically symmetric modes and a corresponding change in
central density.  We deduce that to all orders in $\epsilon$, any
changes in central density must scale in the same way as in Newtonian
theory:
\beq
{\delta \rho_c \over \rho_c} \, \propto \, \alpha^6
\label{final-scaling}
\endeq
as $\alpha \to 0$; see Table \ref{table1} below.  Our result is
restricted by the assumption that the neutron stars are not spinning.
This restriction is unimportant since in Ref.~\cite{ReturnOfTheJedi} the
crushing effect is seen when the neutron stars have very small net
spins.  
Note that the result (\ref{final-scaling}) is not based on an
analysis that includes ``only tidal interactions''.  Instead, our analysis
is a demonstration that there are no general-relativistic interactions that
contribute to the leading-order change in central density 
other than the Newtonian-type tidal interactions.

Our result (\ref{final-scaling}) is in
disagreement with Refs.~\cite{wmm,ReturnOfTheJedi} in the regime
$\alpha \ll 1$ where there should be agreement: Fig.\ 1 of
Ref.~\cite{ReturnOfTheJedi}
shows that the scaling found in the numerical simulations
is $\delta \rho_c / \rho_c \propto \alpha^2$ as $\alpha \to 0$.  This
disagreement in scaling strongly suggests that the star
crushing effect seen in Refs.\ \cite{wmm,ReturnOfTheJedi} is not
physical.
% \cite{note111}.

Our derivation of the result (\ref{final-scaling}) takes place in two
stages.  First, in Secs.\ \ref{sec:main}, \ref{sec:internal-scheme} and
\ref{sec:response}, we analyze the change in central density for
a star moving in a fixed, external, vacuum gravitational field.  We
show that in this context $\delta \rho_c$ scales as ${\cal R}^{-4}$,
where ${\cal R}$
is the radius of curvature of the external spacetime [cf.\ Eq.\
(\ref{finalans0}) below].  In Sec.\ \ref{sec:PN} we extend the
analysis to two stars in a binary, and deduce the scaling
(\ref{final-scaling}).

We use units in which the speed of light $c$ and Newton's gravitational
constant $G$ are unity, and use the sign conventions of Ref.\ \cite{MTW}.
Indices $a,b,c, \ldots$ will be abstract spacetime indices in the
sense of Wald \cite{Waldbook}, thus equations involving such indices
will be valid in all coordinate systems.  Indices
$\alpha$, $\beta$, $\gamma, \ldots$ and  $i,j,k,
\ldots$ will be conventional indices, the former running over
$0,1,2,3$, the latter running over $1,2,3$.

\section{INTERACTIONS OF A FREELY FALLING BODY WITH AN EXTERNAL
GRAVITATIONAL FIELD}
\label{sec:main}

In this section and in the following two sections we shall consider a
neutron star moving in
some arbitrary background vacuum gravitational field, for example a
supermassive black hole.  The key technical tool in our
analysis is the method of {\it matched asymptotic expansions}, as
explained in, for example, Refs.~\cite{Burke,TH,Mino}.  This method
has been used in the past in general relativity primarily to derive
equations of motion for bodies moving in external gravitational fields
\cite{TH,Mino}.  However, the method also lends itself naturally to
analyzing the effect of an external gravitational field on the
internal structure of a fully relativistic, self-gravitating body.
The key feature of the matched asymptotic expansion method is a
separation of lengthscales/timescales: the radius of curvature of the
external spacetime is assumed to be much larger than the lengthscales
characterizing the neutron star, and the timescales over which the
external curvature is changing (as perceived on the neutron star's
worldline) is much longer than the internal dynamical timescale of the
neutron star.

The methods and result which we discuss in this section and in Secs.\
\ref{sec:internal-scheme} and \ref{sec:response} below
do not apply directly
to two stars of comparable mass in a binary, since in that context
there is no external, fixed, background gravitational field.  In Sec.\
\ref{sec:PN} we show how to mesh the matched asymptotic
expansion method with the post-Newtonian expansion method, and extend
our analysis to be applicable to two stars in a binary.
The discussion of motion in fixed external gravitational field of
Secs.\ \ref{sec:main} -- \ref{sec:response} is included as background
and motivation for the analysis of Sec. \ref{sec:PN}.

\subsection{Constructing the spacetime: setup}
\label{main:setup}

Let the metric of the background field (for example, the supermassive
black hole) be $g_{ab}^{(B)}$, and suppose that there are no matter sources
in the region of
interest in the background spacetime, i.e., it is a solution of the
vacuum Einstein equations.  Suppose also that $\Gamma$ is some
geodesic in this spacetime.  Then, one can pick Fermi-normal
coordinates ${\bar x}^\alpha = ({\bar t},{\bar x}^i)$
adapted to this geodesic, such that $\Gamma$ is the curve ${\bar
x}^i=0$, and such that the metric is of the form
\begin{equation}
g_{ab}^{(B)} = {\bar \eta}_{ab} + h^{(B)}_{ab}
\label{background0a}
\end{equation}
where ${\bar \eta}_{ab}$ is the flat, Minkowski metric
${\bar \eta}_{ab} = - (d {\bar t})_a (d {\bar t})_b +
\delta_{ij} (d{\bar x}^i)_a (d {\bar x}^j)_b$, and
where $h^{(B)}_{ab}$ is quadratic in distance from the geodesic.  More
specifically, near the geodesic $\Gamma$ we have \cite{FW}
\begin{eqnarray}
h^{(B)}_{\alpha\beta} d{\bar x}^\alpha d{\bar x}^\beta &=& -R_{0l0m}
{\bar x}^l {\bar
x}^m \, d{\bar t}^2 -
{4 \over 3}
R_{0ljm} {\bar x}^l {\bar x}^m \, d{\bar t} d{\bar x}^j \nonumber \\
\mbox{} && - {1 \over 3} R_{iljm} {\bar x}^l {\bar x}^m
\, d{\bar x}^i d{\bar x}^j \, + O(|{\bar x}^i|^3),
\label{background0}
\end{eqnarray}
where the various Riemann tensor components are evaluated at $({\bar
t},{\bar x}^i)
= ({\bar t},0)$, i.e., on $\Gamma$.

We next introduce some notation.  Let ${\cal R}_c$ denote the order of
magnitude of the radii of curvature
of the background spacetime along $\Gamma$, so that ${\cal R}_c^{-2} = $
typical value of the Riemann tensor components on the right hand side of
Eq.~(\ref{background0}).  Let ${\cal L}$ and ${\cal T}$ be the
lengthscale and timescale, respectively, over which the curvature is changing,
given schematically by $R_{\alpha\beta\gamma\delta} /{\cal L} \sim
\nabla_i R_{\alpha\beta\gamma\delta}$ and
$R_{\alpha\beta\gamma\delta} /{\cal T} \sim \nabla_0
R_{\alpha\beta\gamma\delta}$.  Below we will treat
${\cal R}_c$, ${\cal L}$ and ${\cal T}$ as formally of the same
magnitude, and will denote these lengthscales collectively by ${\cal
R}$.

Consider now a completely different spacetime: a static, spherical,
isolated neutron star, which we model as
a perfect fluid obeying some equation of state $p = p(\rho)$ \cite{extra}.
Let its mass be $M$ and its Schwarschild radius be $R$.
In a suitable coordinate system $x^\alpha = (t,x^i)$,
the neutron star metric can be written as
\begin{equation}
g_{ab}^{({\rm NS})} = \eta_{ab} + h^{({\rm NS})}_{ab},
\label{neutronstar0}
\end{equation}
where ${\eta}_{ab} = - (dt)_a (dt)_b + \delta_{ij} (dx^i)_a (dx^j)_b$,
$r = \sqrt{\delta_{ij} x^i x^j}$, and at large $r$
\FL
\begin{eqnarray}
h^{({\rm NS})}_{\alpha\beta} dx^\alpha dx^\beta &=& -\left[-2 M/r + 2
M^2 / r^2 \right] \,
dt^2 \nonumber \\
\mbox{} && + \left[ {2 M \over r} + {3 M^2 \over 2 r^2}
\right] \delta_{jk} dx^j dx^k \nonumber \\
\mbox{} &&
+ O\left({1 \over r^3}\right).
\label{neutronstar1}
\end{eqnarray}

The task now is to construct, starting from the metrics
(\ref{background0a}) and (\ref{neutronstar0}), an approximate solution
of Einstein's equations that represents the neutron star traveling
along the curve $\Gamma$ in the background spacetime,
in the limit where
$
R \ll {\cal R}.
$
In Sec.\ \ref{main:leadingorder} we describe, without proof, the resulting
spacetime to lowest order in the dimensionless parameters
\begin{equation}
\gamma_1 = R / {\cal R}
\label{gammadef}
\end{equation}
and
\begin{equation}
\gamma_2 = M / {\cal R}.
\label{gammadef2}
\end{equation}
This leading order spacetime is well known.  Then, starting in Sec.\
\ref{main:generalmethod} we describe a systematic
procedure for calculating the spacetime metric to successive orders
in the parameters (\ref{gammadef}) and (\ref{gammadef2})
\cite{TH,Mino}, which can be used to verify the results of Sec.\
\ref{main:leadingorder}.   Finally in Sec.\ \ref{sec:response} we deduce
the scaling of the change in central density of the star.

\subsection{The leading order spacetime}
\label{main:leadingorder}

Consider the region $R \ll r \ll {\cal R}$ in
the background spacetime (\ref{background0}), which we will call the
matching region. In this region, consider the metric
\begin{equation}
g_{ab}^{({\rm MATCHING})} = \eta_{ab} + h_{ab}^{({\rm NS})} + h_{ab}^{(B)}
\label{fullsimple}
\end{equation}
obtained by identifying the coordinates used in Eqs.~(\ref{background0})
and (\ref{neutronstar1}) and by simply adding the metric
perturbations.  The superscript (MATCHING) indicates the region of
spacetime in which the expression is valid.
The metric (\ref{fullsimple}) is an approximate solution of the vacuum
Einstein equation in the matching region, since $h_{ab}^{({\rm NS})}$ and
$h_{ab}^{(B)}$ are both approximate solutions of the linearized vacuum
Einstein equation.  To leading order in the
dimensionless expansion parameters (\ref{gammadef}) and (\ref{gammadef2}),
the metric (\ref{fullsimple}) is the correct, physical metric in the
matching region.

What of outside the matching region?  The physical metric of the
spacetime, again to leading order in the tidal expansion parameters,
can be obtained as follows.  In the region $r \sim R$, which we will
call the interior region, the metric is
\begin{equation}
g_{ab}^{({\rm INTERIOR})} = \eta_{ab} + h_{ab}^{({\rm NS})} + {\hat
h}_{ab}^{(B)}.
\label{interiormetric}
\end{equation}
Here, the tensor $h_{ab}^{({\rm NS})}$ is the full,
nonlinear metric perturbation of the neutron star.
The tensor ${\hat h}_{ab}^{(B)}$ is a linearized metric perturbation
which describes the leading order effect of the external tidal field.
This metric perturbation, together with some linearized perturbation to the
star's fluid variables, is a solution of the coupled Einstein plus
perfect fluid equations, linearized about the neutron star background,
and with all time derivative terms dropped \cite{note43}.  [The time
derivative terms
scale as $1/{\cal T} \sim 1/ {\cal R}$ and thus enter at higher order
in the small parameters (\ref{gammadef}) and (\ref{gammadef2}); see
Sec.~\ref{main:generalmethod} below and Ref.\ \cite{Burke}.]
The solution ${\hat h}_{ab}^{(B)}$ is chosen such that, at large $r$,
${\hat h}_{ab}^{(B)} \to $ the quadratic expression on the right hand
side of Eq.~(\ref{background0}).
These perturbations (of the metric and of the fluid
variables) describe normal mode deformations \cite{note88} of the
star responding adiabatically to the external tidal field.
The boundary condition at large $r$ determines both the metric
perturbation ${\hat h}_{ab}^{(B)}$ and the perturbations to
the fluid variables.  The situation is analogous to that in Newtonian
gravity as analyzed in detail in Ref.~\cite{Dong1}, where the boundary
condition on the Newtonian potential at large $r$ determines the
solution of the stellar perturbation equations.

In a similar way, the spacetime metric in the exterior region $r \agt
{\cal R}$ can be written as
\begin{equation}
g_{ab}^{({\rm EXTERIOR})} = {\bar \eta}_{ab} + {\hat h}_{ab}^{({\rm NS})} +
h_{ab}^{(B)}.
\label{exteriormetric}
\end{equation}
Here, $h_{ab}^{(B)}$ is of order unity, and ${\hat h}_{ab}^{({\rm NS})}$ is
a linearized solution of the vacuum Einstein equations on the
background $g_{ab}^{(B)}$ in the exterior region, which matches
smoothly onto $h_{ab}^{({\rm NS})}$ in the matching region.  To
leading order, ${\hat h}_{ab}^{({\rm NS})}$ is just the solution of
the linearized vacuum Einstein equations on the background
$g_{ab}^{(B)}$ whose source is a delta function along the worldline
$\Gamma$.

\subsection{Constructing the spacetime: general method}
\label{main:generalmethod}

In this section we describe a general scheme for calculating the
spacetime metric perturbatively in the parameters (\ref{gammadef})
and (\ref{gammadef2}), based on the treatment in Thorne and Hartle
\cite{TH} and in Mino et.\ al. \cite{Mino}.  This general scheme
justifies the
claimed forms (\ref{interiormetric}) and (\ref{exteriormetric}) above
of the leading order metrics.
It also enables us to go to higher order, which
is necessary since (as in Newtonian and post-Newtonian theory) it
turns out that changes in the central density of the star are
quadratic in the leading order tidal field ${\hat
h}_{ab}^{(B)}$.  Therefore,
in order to determine the leading order change in central density
of the star, it is necessary to consider higher order perturbations
which scale as $({\hat h}_{ab}^{(B)})^2$.

There are three elements to the general procedure: an internal
scheme, an external scheme, and a matching of the two schemes \cite{Mino}.

\subsubsection{The internal scheme}
\label{internalscheme}

Let $(M^{({\rm NS})},g_{ab}^{({\rm NS})})$ denote the manifold and
background metric (\ref{neutronstar0}) of the neutron star.  The
physical metric can be written as the background metric plus a
sequence of perturbations of various orders, in a generalization of
Eq.~(\ref{interiormetric}):
\begin{eqnarray}
g_{ab}^{({\rm INTERIOR})} &=& g_{ab}^{({\rm NS})} +  \varepsilon
{\hat h}_{ab}^{(1)}
+  \varepsilon^2 {\hat h}_{ab}^{(2)}
+  \varepsilon^3 {\hat h}_{ab}^{(3)}
\nonumber \\
\mbox{} &&
+  \varepsilon^4 {\hat h}_{ab}^{(4)} + O(\varepsilon^5).
\label{interiormetric1}
\end{eqnarray}
Here ${\hat h}_{ab}^{(2)}$ is what we called ${\hat h}_{ab}^{(B)}$
above, and below we will show that ${\hat h}_{ab}^{(1)}=0$.
The quantity $\varepsilon$ is a formal expansion parameter which can
be set to one at the end of the calculation; each term $\varepsilon^t
{\hat h}_{ab}^{(t)}$ scales like ${\cal R}^{-t}$ (but has no definite
scaling with respect to $M$).  We shall be working to order
$O(\varepsilon^4)$.   The expansion (\ref{interiormetric1})
constitutes a general-relativistic generalization of the usual tidal
expansion of stellar interactions in Newtonian gravity.
In a similar way we can expand the stress-energy tensor $T_{ab}$ of the
neutron star as
\begin{equation}
T_{ab} = T_{ab}^{(0)} + \varepsilon T_{ab}^{(1)} + \varepsilon^2
T_{ab}^{(2)} + \varepsilon^3 T_{ab}^{(3)} + \varepsilon^4 T_{ab}^{(4)}
+ \ldots.
\label{stresstensor1}
\end{equation}
Here $T_{ab}^{(0)}$ is the stress-tensor of the static, spherical,
unperturbed neutron star, and the higher order terms $\varepsilon^t
T_{ab}^{(t)}$ scale as ${\cal R}^{-t}$.

The terms with $t=1$ in Eq.\ (\ref{interiormetric1}) and
(\ref{stresstensor1}) actually vanish identically.
This is because when one carries out the matching procedure described
below to determine the solutions, one finds that 
there are no pieces
of the internal solution that scale as ${\cal R}^{-1}$ \cite{TH}.
Henceforth we shall anticipate this result and set ${\hat
h}_{ab}^{(1)} = T_{ab}^{(1)} = 0$.

Now the perturbations to the interior metric and to the neutron star
are driven by the external gravitational fields which vary over a time
scale ${\cal T} \sim {\cal R} \,\propto\, 1/\varepsilon$.  This fact needs to
be built into the
approximation scheme we use to derive the perturbation equations of
motion.  We can illustrate the nature of the required approximation
scheme with the following example.  Consider
a scalar field $\Phi$ on flat spacetime obeying the equation
\beq
\Box \Phi = \left(-{\partial^2 \over \partial t^2} + {\bf \nabla}^2
\right) \Phi({\bf x},t) = \rho({\bf x},t),
\label{example-scaling}
\endeq
where the source $\rho({\bf x},t)$ is of the form $\rho({\bf x},t) =
\rho_0({\bf x},\varepsilon t)$, and the function $\rho_0$ is
independent of $\varepsilon$.  If we define $\Phi_0({\bf x},t) =
\Phi({\bf x},t/\varepsilon)$
and $\tau = \varepsilon t$, Eq.\ (\ref{example-scaling}) can be
re-written as
\beq
\left(-\varepsilon^2 {\partial^2 \over \partial \tau^2} + {\bf \nabla}^2
\right) \Phi_0({\bf x},\tau) = \rho_0({\bf x},\tau).
\label{example-scaling1}
\endeq
If $\rho_0$ is now specified as a power series in $\varepsilon$, we can
solve for $\Phi_0$ order by order in $\varepsilon$ using Eq.\
(\ref{example-scaling1}).
The equation of motion (\ref{example-scaling1}) can also be obtained
(dropping the subscripts 0) simply by multiplying each time derivative in Eq.\
(\ref{example-scaling}) by $\varepsilon$.

In a similar way, the equations of motion of the
internal scheme can be obtained by writing out the Einstein and fluid
equations in terms of the contravariant components $g_{ab}$ and
$T_{ab}$ of the metric and stress tensor, substituting in the
expansions (\ref{interiormetric1}) and (\ref{stresstensor1}), and by
multiplying each time derivative by $\varepsilon$.
More specifically, for each $\varepsilon$, there will exist an
$\varepsilon$-dependent
coordinate system $({\tilde x}^i, {\tilde t})$ which varies smoothly
in $\varepsilon$ and which coincides with the coordinate system
$(x^i,t)$ of Eq.\ (\ref{neutronstar1}) at $\varepsilon=0$, such that
\beq
{\hat h}^{(t)}_{\alpha\beta}({\tilde x}^i,{\tilde t}) d {\tilde
x}^\alpha d {\tilde x}^\beta =
{\tilde h}^{(t)}_{\alpha\beta}({\tilde x}^i,\varepsilon {\tilde t}) d {\tilde
x}^\alpha d {\tilde x}^\beta
\label{rescale-gab}
\endeq
and
\beq
T^{(t)}_{\alpha\beta}({\tilde x}^i,{\tilde t}) d {\tilde x}^\alpha d {\tilde
x}^\beta =
{\tilde T}^{(t)}_{\alpha\beta}({\tilde x}^i,\varepsilon {\tilde t}) d {\tilde
x}^\alpha d {\tilde x}^\beta
\label{rescale-Tab}
\endeq
for $t = 1,2,3 \ldots$.  Here the tensors ${\tilde h}^{(t)}_{ab}$
and ${\tilde T}^{(t)}_{ab}$ are independent of $\varepsilon$.
The Einstein and perfect fluid equations
combined with Eqs.\
(\ref{interiormetric1}), (\ref{stresstensor1}), (\ref{rescale-gab})
and (\ref{rescale-Tab}) now yield a system of equations for the
tensors ${\tilde h}^{(t)}_{ab}$ and ${\tilde T}^{(t)}_{ab}$ analogous
to Eq.\ (\ref{example-scaling1}).  These equations will only be valid
in the restricted class of coordinate systems for which the ansatz
(\ref{rescale-gab}) and (\ref{rescale-Tab}) are valid.

The approximation scheme can also be described in
coordinate invariant terms in the following way.
Let ${\tilde t}(\varepsilon)$ be a scalar field and
${\tilde t}^a(\varepsilon)$ a
vector field with
\beq
{\tilde t}^a \nabla_a {\tilde t}=1,
\label{t-normalization}
\endeq
which reduce
to the $t$ and $\partial / \partial t$ of the coordinate system
(\ref{neutronstar1}) at $\varepsilon=0$.
Define the tensor
\beq
\Lambda_a^b(\varepsilon) = \delta_a^b + \left({1 \over
\varepsilon}-1\right) {\tilde t}^b \nabla_a {\tilde t}
\label{Lambdadef}
\endeq
Let $\chi_\varepsilon: M^{({\rm NS})} \to M^{({\rm NS})}$ be the
mapping which moves any point $\ln \varepsilon$ units along integral
curves of the vector field ${\tilde t} \, {\tilde t}^a$, i.e., in
suitable coordinate systems $({\tilde x}^i,{\tilde t})$,
$\chi_\varepsilon$ maps $({\tilde x}^i,{\tilde t})$ to $({\tilde
x}^i,\varepsilon {\tilde t})$.
We make the following ansatz for the form of the metric
$g_{ab}^{({\rm INTERIOR})}$ and the stress tensor $T_{ab}$:
\beq
g_{ab}^{({\rm INTERIOR})}(\varepsilon) = \Lambda_a^c(\varepsilon)
\Lambda_b^d(\varepsilon) \, \chi_{\varepsilon\,*} {\tilde g}_{cd},
\label{ansatz1}
\endeq
and
\beq
T_{ab}(\varepsilon) = \Lambda_a^c(\varepsilon)
\Lambda_b^d(\varepsilon) \, \chi_{\varepsilon_*} {\tilde T}_{cd},
\label{ansatz2}
\endeq
where
\begin{eqnarray}
{\tilde g}_{ab} &=& g_{ab}^{({\rm NS})} +  \varepsilon
{\tilde h}_{ab}^{(1)}
+  \varepsilon^2 {\tilde h}_{ab}^{(2)}
+  \varepsilon^3 {\tilde h}_{ab}^{(3)}
\nonumber \\
\mbox{} &&
+  \varepsilon^4 {\tilde h}_{ab}^{(4)} + O(\varepsilon^5)
\label{interiormetric2}
\end{eqnarray}
and
\begin{equation}
{\tilde T}_{ab} = T_{ab}^{(0)} + \varepsilon {\tilde T}_{ab}^{(1)} +
\varepsilon^2
{\tilde T}_{ab}^{(2)} + \varepsilon^3 {\tilde T}_{ab}^{(3)} +
\varepsilon^4 {\tilde T}_{ab}^{(4)}
+ O(\varepsilon^5).
\label{stresstensor2}
\end{equation}
Here the tensors ${\tilde h}_{ab}^{(t)}$ and ${\tilde T}_{ab}^{(t)}$
are independent of $\varepsilon$ and $\chi_{\varepsilon\,*}$ is
the pullback map.  Equations (\ref{ansatz1}) -- (\ref{stresstensor2})
together with Eqs.\ (\ref{interiormetric1}) -- (\ref{stresstensor1})
reduce to Eqs.\ (\ref{rescale-gab}) and (\ref{rescale-Tab}) in
suitable coordinate systems.  [Note that the transformation in Eqs.\
(\ref{ansatz1}) and (\ref{ansatz2}) leave the background quantities
$g_{ab}^{({\rm NS})}$ and $T_{ab}^{(0)}$ invariant.]
Finally, the Einstein and perfect fluid equations
\begin{eqnarray}
G_{ab}[g^{({\rm INTERIOR})}_{cd}] &=& 0
\label{einstein}
\\
\mbox{} \nabla^a T_{ab} &=&0,
\label{pfluid}
\end{eqnarray}
when expanded order by order in $\varepsilon$, yield a system of
elliptic equations for the tensors ${\tilde h}_{ab}^{(t)}$ and
${\tilde T}_{ab}^{(t)}$.

Below, we shall for simplicity set $\varepsilon=1$ after having
derived the equations of motion.  When $\varepsilon=1$ we have
${\tilde h}_{ab}^{(t)} = {\hat h}_{ab}^{(t)}$ and ${\tilde
T}_{ab}^{(t)} = T_{ab}^{(t)}$, and thus we can write the
equations in terms of the fields ${\hat h}_{ab}^{(t)}$ and
$T_{ab}^{(t)}$.

In order to write out explicitly the resulting equations of motion, we
introduce the following notations.  We define the operators
$G_{ab}^{(1)}[h; g_{cd}^{({\rm
NS})}]$ and $G_{ab}^{(2)}\left[h,h; g_{ab}^{({\rm NS})}\right]$, which
act on metric perturbations and pairs of metric perturbations
respectively, to be linear and quadratic parts of the Einstein tensor
of the perturbed spacetime $g^{({\rm NS})} + \varepsilon h$:
\begin{eqnarray}
G_{ab}[g^{({\rm NS})} + \varepsilon h] &=&
G_{ab}[g^{({\rm NS})}] +
\varepsilon
G^{(1)}_{ab}[h;g^{({\rm NS})}]
\nonumber \\ \mbox{} &&
+ \varepsilon^2 G^{(2)}_{ab}[h,h; g^{({\rm NS})}]
+ O(\varepsilon^3).
\label{G12def}
\end{eqnarray}
[Here and below we drop some of the tensorial indices for ease of
notation].
We also use the expansions
\begin{equation}
G_{ab}^{(1)}[h] = G_{ab}^{(1,0)}[h] + G_{ab}^{(1,1)}[h]
+ G_{ab}^{(1,2)}[h]
\end{equation}
and
\begin{equation}
\FL
G_{ab}^{(2)}[h,h] = G_{ab}^{(2,0)}[h,h] + G_{ab}^{(2,1)}[h,h]
+ G_{ab}^{(2,2)}[h,h],
\end{equation}
where $G_{ab}^{(1,p)}$ and $G_{ab}^{(2,p)}$ scale as
${\cal T}^{-p}$, {\it i.e.}, contain $p$ time derivatives, for $p = 0,1,2$.
We write the derivative operator for the metric $g_{ab}^{({\rm NS})}
+ \varepsilon h_{ab}$, for any perturbation $h_{ab}$, as
\begin{equation}
\nabla^a = \nabla^{(0)\,a} + \varepsilon \nabla^{(1)\,a}_{[h]} +
\varepsilon^2 \nabla^{(2)\,a}_{[h,h]} + O(\varepsilon^3).
\label{expand1}
\end{equation}
This equation defines the operators $\nabla^{(1)\,a}_{[h]}$ and
$\nabla^{(2)\,a}_{[h,h]}$, which are multiplicative and not
differential operators despite the notation.
Finally, we use the expansion
\begin{equation}
\nabla^{(t)\,a} = \sum_{p=0} \nabla^{(t,p)\,a},
\label{expand2}
\end{equation}
for $t = 0,1,2$, where  $\nabla^{(t,p)\,a}$ contains $p$ time
derivatives (either
acting on $h$ or as differential operators) and thus scales as ${\cal
T}^{-p}$.

Using these notations and the expansions (\ref{interiormetric1}) and
(\ref{stresstensor1}), we obtain from Eq.\ (\ref{pfluid})
the perturbed fluid equations \cite{extra}
\begin{equation}
\nabla^{(0,0)\,a} T_{ab}^{(2)} + \nabla^{(1,0)\,a}_{[{\hat
h}^{(2)}]} T_{ab}^{(0)} = 0,
\label{fluid2}
\end{equation}
\begin{equation}
\nabla^{(0,0)\,a} T_{ab}^{(3)} + \nabla^{(1,0)\,a}_{[{\hat
h}^{(3)}]} T_{ab}^{(0)} = {\cal F}^{(3)}_b,
\label{fluid3}
\end{equation}
and
\begin{equation}
\nabla^{(0,0)\,a} T_{ab}^{(4)} + \nabla^{(1,0)\,a}_{[{\hat h}^{(4)}]}
T_{ab}^{(0)} = {\cal F}^{(4)}_b,
\label{fluid4}
\end{equation}
where the 4-force densities ${\cal F}^{(3)}_b$ and ${\cal F}^{(4)}_b$ are
\begin{equation}
{\cal F}^{(3)}_b = - \nabla^{(1,1)\,a}_{[{\hat h}^{(2)}]}
T_{ab}^{(0)}  - \nabla^{(0,1)\,a} T_{ab}^{(2)}
\label{forcedef}
\end{equation}
and
\begin{eqnarray}
{\cal F}^{(4)}_b &=& - \nabla^{(1,2)\,a}_{[{\hat h}^{(2)}]} T_{ab}^{(0)}
- \nabla^{(1,1)\,a}_{[{\hat h}^{(3)}]} T_{ab}^{(0)}
 - \nabla^{(0,1)\,a} T_{ab}^{(3)} \nonumber \\
\mbox{} &&
- \nabla^{(2,0)\,a}_{[{\hat h}^{(2)},{\hat h}^{(2)}]} T_{ab}^{(0)}
-  \nabla^{(1,0)\,a}_{[{\hat h}^{(2)}]} T_{ab}^{(2)}.
\label{forcedef1}
\end{eqnarray}
Note that the actual fluid equations of motion are obtained by using the
formula $T_{ab} = (\rho+p) u_a u_b + p g_{ab}$, by assuming
expansions of the form (\ref{stresstensor1}) for the density $\rho$
and four velocity $u^a$, and by substituting into Eqs.\ (\ref{fluid2}) --
(\ref{fluid4}).  However, the schematic form (\ref{fluid2}) --
(\ref{fluid4}) will be sufficient for our purposes.

Similarly, from the Einstein equations we obtain the equations
of motion for the metric perturbations ${\hat h}_{ab}^{(t)}$ for $m =
2,3,4$ :
\begin{equation}
G_{ab}^{(1,0)}\left[ {\hat h}^{(2)}; g_{cd}^{({\rm NS})}\right] =
8 \pi T_{ab}^{(2)},
\label{hB2}
\end{equation}
\begin{equation}
G_{ab}^{(1,0)}\left[ {\hat h}^{(3)}; g_{cd}^{({\rm NS})}\right] =
8 \pi T_{ab}^{(3)} - G_{ab}^{(1,1)}\left[ {\hat h}^{(2)};
g_{cd}^{({\rm NS})}\right],
\label{hB3}
\end{equation}
and
\begin{eqnarray}
\FL
G_{ab}^{(1,0)}\left[ {\hat h}^{(4)}; g_{cd}^{({\rm NS})}\right] &=&
8 \pi T_{ab}^{(4)}
- G_{ab}^{(1,1)}\left[ {\hat h}^{(3)};g_{cd}^{({\rm NS})}\right]
 \nonumber \\ \mbox{} &&
- G_{ab}^{(2,0)}\left[{\hat h}^{(2)},{\hat h}^{(2)}; g_{cd}^{({\rm
NS})}\right] \nonumber \\
\mbox{} &&
- G_{ab}^{(1,2)}\left[ {\hat h}^{(2)};g_{cd}^{({\rm NS})}\right].
\label{hB4}
\end{eqnarray}
Note that Eqs.~(\ref{fluid2}) -- (\ref{fluid4}) and also Eqs.\
(\ref{hB2}) -- (\ref{hB4}) are elliptic and not hyperbolic.

As discussed in Sec.~\ref{main:leadingorder}, Eqs.\ (\ref{fluid2}) and
(\ref{hB2}) for the leading order fields ${\hat h}_{ab}^{(2)}$ and
$T_{ab}^{(2)}$ together describe a solution of the
Einstein plus perfect fluid equations, linearized about the static
neutron star background, with all time derivative terms dropped.
The higher order equations (\ref{fluid2}), (\ref{fluid3}), (\ref{hB3})
and (\ref{hB4}) have the same basic structure, but contain additional
source terms.  We shall be interested in very general solutions of the
equations which are allowed to diverge as $r \to \infty$; the physical
solutions will be determined by matching to the metric of the external
scheme.

\subsubsection{The external scheme}
\label{externalscheme}

Let $(M^{({\rm B})},g_{ab}^{({\rm B})})$ denote the manifold and
metric (\ref{background0a}) of the external spacetime.  The
physical metric can be written as this background metric plus a
sequence of perturbations of various orders, in a generalization of
Eq.~(\ref{exteriormetric}):
\begin{eqnarray}
g_{ab}^{({\rm EXTERIOR})} &=& g_{ab}^{({\rm B})} +  \varepsilon {\bar
h}_{ab}^{(1)}
+  \varepsilon^2 {\bar h}_{ab}^{(2)} + O(\varepsilon^3).
\label{exteriormetric1}
\end{eqnarray}
Here as above $\varepsilon$ is a formal expansion parameter which can
be set to one at the end of the calculation; in
Eq.~(\ref{exteriormetric1}), each term $\varepsilon^s
{\bar h}_{ab}^{(s)}$ scales like $M^{s}$ but has no definite
scaling with respect to ${\cal R}$.  [Here we treat the quantities $M$
and $R$ as formally of the same magnitude, so that the expansion
parameters (\ref{gammadef}) and (\ref{gammadef2}) coincide \cite{note6}].
The perturbation ${\bar
h}_{ab}^{(1)}$ was denoted above as ${\hat h}_{ab}^{({\rm NS})}$.
The equations of motion satisfied by the metric perturbations ${\bar
h}_{ab}^{(1)}$ and ${\bar h}_{ab}^{(2)}$ are obtained from the vacuum
Einstein equation:
\begin{eqnarray}
\FL
G_{ab}^{(1)}\left[ {\bar h}^{(1)}; g_{cd}^{({\rm B})}\right] &=& 0
\label{externaleqns1}
\\ \mbox{}
G_{ab}^{(1)}\left[ {\bar h}^{(2)}; g_{cd}^{({\rm B})}\right] +
G_{ab}^{(2)}\left[{\bar h}^{(1)},{\bar h}^{(1)}; g_{cd}^{({\rm
B})}\right] &=& 0.
\label{externaleqns}
\end{eqnarray}
The solutions to these equations will diverge as ${\bar r} \to 0$, and
will be determined by matching onto the internal solutions.

\subsubsection{The matching scheme}
\label{matchingscheme}

In the matching region $R \ll r \ll {\cal R}$, the interior metric
(\ref{interiormetric1}) can be written as the double expansion
\cite{note6}:
\begin{equation}
g_{ab}^{({\rm INTERIOR})} = \eta_{ab} + \sum_{s=0} \,
\sum_{t=0} \, {\hat h}_{ab}^{(s,t)} \, \left( {M \over r} \right)^s \,
\left( {r \over {\cal R}} \right)^t.
\label{doubleexpansion}
\end{equation}
Here the terms with $t=0$ are the expansion in powers of
$M/r$ of the neutron star metric perturbation $h_{ab}^{({\rm NS})}$,
and the terms with $t=1$ are the expansion in powers of $M/r$ of the
metric perturbation ${\hat h}_{ab}^{(1)}$.
The terms with $t=2$ give the expansion in powers of $M/r$ of the
metric ${\hat h}_{ab}^{(2)} = {\hat h}_{ab}^{(B)}$, which describes
the leading order tidal gravitational field, and so forth.
The form of the
expansion (\ref{doubleexpansion}) can be regarded as an ansatz which
is validated by explicit matching calculations to the first few orders
in $s$ and $t$ \cite{Mino}; however, at higher orders in $s$ and $t$
it might be necessary to include powers of $ (M/r) \log[M/r]$ or
$(r/{\cal R}) \log[r/{\cal R}]$ \cite{notelog}.  Moreover, the
expansion is merely
asymptotic and is not expected to be convergent \cite{Burke}.
{}From Eqs.~(\ref{hB2}) --
(\ref{hB4}), each term $h_{ab}^{(s,t)} (M/r)^s (r / {\cal
R})^t$ in Eq.~(\ref{doubleexpansion}) obeys a linear elliptic equation
with nonlinear source terms generated by the terms $h_{ab}^{(s',t')}
(M/r)^{s'} (r / {\cal
R})^{t'}$ with $s' < s$ and/or $t' < t$.  The solutions for $t \ge 1$
diverge as $r \to \infty$.

In a similar way, we can perform a double expansion of the external
metric (\ref{exteriormetric1}):
\begin{equation}
g_{ab}^{({\rm EXTERIOR})} = {\bar \eta}_{ab} + \sum_{s=0} \,
\sum_{t=0} \, {\bar h}_{ab}^{(s,t)} \, \left( {M \over {\bar r}} \right)^s \,
\left( {{\bar r} \over {\cal R}} \right)^t,
\label{doubleexpansion1}
\end{equation}
where ${\bar r}^2 = \delta_{ij} {\bar x}^i {\bar x}^j$.  Here the
terms of a given order $s$ give the expansion in powers of ${\bar r}
/{\cal R}$ of the metric perturbation ${\bar h}_{ab}^{(s)}$, while the
terms with $s=0$ are the expansion of the background metric
$g_{ab}^{(B)}$.

The basic idea now is to demand consistency between the expansions
(\ref{doubleexpansion}) and (\ref{doubleexpansion1}).  Before this can
be done, however, one must specify an embedding
\begin{equation}
\varphi: \ M^{({\rm NS})} \to M^{({\rm B})},
\end{equation}
of the neutron star spacetime into the external spacetime, i.e.,
a mapping between the asymptotically Lorentzian
coordinates $x^\alpha$ of $M^{({\rm NS})}$ and the Fermi-normal coordinates
${\bar x}^\alpha$ of $M^{({\rm B})}$.
Let us write this mapping as
\begin{eqnarray}
{\bar x}^\alpha&=& {\bar x}^\alpha(t,x^i) \nonumber \\
\mbox{} &=&
\sum_{p=0} \, \left( { M \over {\cal R} } \right)^p \, \sum_{r=0} \,
\sum_{I_r} \ \ {}^{(r,p)}{\cal F}^\alpha_{I_r}(t) \, x^{I_r},
\label{varphidef}
\end{eqnarray}
where $I_r$ denotes the multi-index $(i_1, \ldots, i_r)$ and $x^{I_r} \equiv
x^{i_1} x^{i_2} \ldots x^{i_r}$.  Equation (\ref{varphidef}) is a
Taylor expansion of $\varphi$ in terms both of the spatial coordinates
$x^i$ at each fixed $t$, and also in terms of the parameter $M / {\cal
R}$.  Thus, the worldline of the center of
the neutron star gets mapped onto the worldline
\begin{equation}
{\bar x}^\alpha(t) = {}^{(0,0)}{\cal F}^\alpha(t) + {M \over {\cal R}}
{}^{(0,1)}{\cal F}^\alpha(t) + O[(M/{\cal R})^2].
\label{centralworldline}
\end{equation}
The matching procedure described below can be used to show that the
first term in Eq.\ (\ref{centralworldline}) represents a geodesic in
the background metric $g_{ab}^{(B)}$, and that the second term is the
first order correction to geodesic motion due to radiation reaction
\cite{Mino}.

Using the expansions (\ref{doubleexpansion1}) and (\ref{varphidef}) we
can calculate the pullback $\varphi_* g_{ab}^{({\rm EXTERIOR})}$ of
the external metric to the manifold $M^{({\rm NS})}$, and expand it
as:
\begin{equation}
\varphi_* g_{ab}^{({\rm EXTERIOR})} = \eta_{ab} + \sum_{s=0} \,
\sum_{t=0} \, {\bar h}_{ab}^{\prime\,(s,t)} \, \left( {M \over r} \right)^s \,
\left( {r \over {\cal R}} \right)^t.
\label{doubleexpansion2}
\end{equation}
Here each term ${\bar h}_{ab}^{\prime\,(s,t)}$ can depend on the quantities
${\bar h}_{ab}^{(s^\prime,t^\prime)}$ with $0 \le s^\prime \le s$ and
$0 \le t^\prime \le t$, and also on the embedding functions
${}^{(r,p)} {\cal F}_{I_r}^\alpha(t)$ with $0 \le r \le t+1$ and $0 \le p
\le s$.
For all choices of these embedding functions, the metric
(\ref{doubleexpansion2}) will satisfy the perturbative vacuum
Einstein equation, since it is the pullback of a metric which does
so.

We can now describe, schematically, the matching procedure.  First,
solve for the general solutions for the metric perturbations up to
a given order $m$ and $l$ in both Eqs.\ (\ref{doubleexpansion}) and
(\ref{doubleexpansion1}).  Second, specify the embedding
(\ref{varphidef}) up to the required order in $r$ and $p$,
and calculate the expansion coefficients ${\bar h}_{ab}^{\prime\,(s,t)}$ in
Eq.~(\ref{doubleexpansion2}).  Third, demand
that the expansions (\ref{doubleexpansion}) and
(\ref{doubleexpansion2}) agree, which determines both the metric
perturbation solutions and also the embedding free functions
${}^{(r,p)} {\cal F}^\alpha_{I_r}(t)$, up to some residual gauge freedom.
This matching procedure simultaneously determines the influence of the
external gravitational field on the internal structure of the neutron
star, and also the influence of the neutron star on the external
field.  The matchings for the cases $(s,t) = (0,0), (0,1),
(0,2), (1,0),
(1,1), (1,2)$ have been explicitly worked out by Mino et.\ al.\
\cite{Mino}, in the case where the interior metric describes a black
hole.  In Sec.\ \ref{sec:response} below we will review these matchings and
also justify some of the assumptions made by Mino et.\ al.\ concerning
the monopole and dipole pieces of the fields.  We shall need to
consider in addition the cases $(s,t) = (0,3),(0,4),(1,3),(1,4)$ and
$(2,4)$.

In order to understanding how the matching works in these cases, we
need to understand in detail the nature of the space of solutions in
the internal scheme.  We examine this issue in the next section.

\section{THE INTERNAL SCHEME : STRUCTURE OF THE SOLUTION SPACE}
\label{sec:internal-scheme}

\subsection{Gauge freedom}
\label{internal-scheme:gauge}

We start by discussing the gauge freedom in the perturbation
equations.  This gauge freedom consists of one parameter families
of diffeomorphisms $\psi_\varepsilon : M^{({\rm NS})} \to M^{({NS})}$,
where $\psi_0$
is the identity map, which act on the quantities
(\ref{interiormetric2}) and (\ref{stresstensor2}) via the natural
pull-back action.  We can express such a map to $O(\varepsilon^4)$ as
\FL
\begin{equation}
\psi_\varepsilon = {\cal D}_{\xi^{(1)}}(\varepsilon) \circ {\cal
D}_{\xi^{(2)}}(\varepsilon^2) \circ {\cal
D}_{\xi^{(3)}}(\varepsilon^3) \circ {\cal D}_{\xi^{(4)}}(\varepsilon^4),
\label{gaugefreedom1}
\end{equation}
where $\xi^{{(p)}\,a}$ for $1 \le p \le 4$ are vector fields
and where for any vector field $\tau^a$, ${\cal D}_\tau(\lambda) : \,
M^{({\rm NS})} \to M^{({\rm NS})}$ denotes the one parameter group of
diffeomorphisms generated
by $\tau^a$.  Since all perturbation quantities vanish at first
order in $\varepsilon$, we can without loss of generality assume that
$\xi^{(1)\, a} =0$.

Now suppose that $S(\varepsilon)$ is any one parameter family of
tensor fields on $M^{({\rm NS})}$ (we suppress tensor indices), which
has the expansion
\begin{equation}
S(\varepsilon) = S^{(0)} + \varepsilon^2 S^{(2)} + \varepsilon^3 S^{(3)} +
\varepsilon^4 S^{(4)} + O(\varepsilon^5).
\end{equation}
Then from Eq.~(\ref{gaugefreedom1}) we can calculate the
transformation properties of the expansion coefficients $S^{(2)}$, $S^{(3)}$
etc.  We find \cite{gaugemethod}
\begin{equation}
\psi^*_\varepsilon S(\varepsilon) =
S^{(0)} + \varepsilon^2 {\bar S}^{(2)} + \varepsilon^3 {\bar S}^{(3)} +
\varepsilon^4 {\bar S}^{(4)} + O(\varepsilon^5).
\end{equation}
where
\begin{equation}
{\bar S}^{(2)} = S^{(2)} + {\cal L}_{\xi^{(2)}} S^{(0)},
\label{gauge2}
\end{equation}
\begin{equation}
{\bar S}^{(3)} = S^{(3)} + {\cal L}_{\xi^{(3)}} S^{(0)},
\label{gauge3}
\end{equation}
and
\begin{eqnarray}
{\bar S}^{(4)} &=& S^{(4)} + {\cal L}_{\xi^{(4)}} S^{(0)}  + {1\over2}
{\cal L}_{\xi^{(2)}} {\cal L}_{\xi^{(2)}} S^{(0)} \nonumber \\
\mbox{} && + {\cal L}_{\xi^{(2)}} S^{(2)}.
\label{gauge4}
\end{eqnarray}
Here ${\cal L}$ means the Lie derivative.  The formulae (\ref{gauge2}) ---
(\ref{gauge4}) apply when we take $S(\varepsilon)$ to be any one of the
fields ${\tilde g}_{ab}(\varepsilon)$, ${\tilde T}_{ab}(\varepsilon)$,
$t(\varepsilon)$ and $t^a(\varepsilon)$ defined in Sec.\
\ref{internalscheme}.

We now fix a certain portion of the gauge freedom.  We can expand the
fields $t(\varepsilon)$ and $t^a(\varepsilon)$ as \cite{notelineart}
\beq
t(\varepsilon) = t^{(0)} + \varepsilon^2 t^{(2)} + \varepsilon^3
t^{(3)} + \varepsilon^4 t^{(4)} + O(\varepsilon^5),
\label{texpand}
\endeq
and
\beq
t^a(\varepsilon) = t^{(0)\,a} + \varepsilon^2 t^{(2)\,a} + \varepsilon^3
t^{(3)\,a} + \varepsilon^4 t^{(4)\,a} + O(\varepsilon^5).
\label{taexpand}
\endeq
{}From Eqs.\ (\ref{t-normalization}) and (\ref{gauge2}) --
(\ref{gauge4}) it follows that we can choose a gauge such that
\beq
t^{(2)} = t^{(3)} = t^{(4)} = t^{(2)\,a} = t^{(3)\,a} = t^{(4)\,a}
=0.
\label{gaugechoice}
\endeq
The residual gauge freedom then consists of the transformations
for which
\beq
{\cal L}_{t^{(0)\,a}} \xi^{(p)\,b} = \xi^{(p)\,a} \nabla_a t^{(0)} =0,
\label{spatial-gauge}
\endeq
for $2 \le p \le 4$.  Thus, in the coordinate system $(x^i,t)$ of Eq.\
(\ref{neutronstar1}), the vector fields are purely spatial,
$\xi^{(p)\,t} =0$ and $\xi^{(p)\,i}(x^j,t) =  \xi^{(p)\,i}(x^j)$.

\subsection{The Newtonian case}
\label{internal-scheme:Newtonian}

To motivate our analyses below, consider first the analogous
Newtonian perturbation theory of a static star \cite{restriction}.
The density is given
by $\rho = \rho^{(0)} + \varepsilon^2 \rho^{(2)} + \varepsilon^3
\rho^{(3)} + \varepsilon^4 \rho^{(4)} + O(\varepsilon^5)$, the
pressure by $p = p^{(0)} + \varepsilon^2 p^{(2)} + \varepsilon^3
p^{(3)} + \varepsilon^4 p^{(4)} + O(\varepsilon^5)$, the velocity by
${\bf u} = \varepsilon^3 {\bf u}^{(3)} + \varepsilon^4 {\bf u}^{(4)}
+ O(\varepsilon^5)$, and the Newtonian potential by $\Phi = \Phi^{(0)}
+ \varepsilon^2 \Phi^{(2)} + \varepsilon^3 \Phi^{(3)} +
\varepsilon^4 \Phi^{(4)} + O(\varepsilon^5)$.  The perturbed
equations of motion [the Newtonian limit of Eqs.\ (\ref{fluid2}) --
(\ref{hB4})] are at $O(\varepsilon^2)$
\FL
\begin{eqnarray}
\nabla^2 \Phi^{(2)} - 4 \pi G \rho^{(2)} &=& 0
\label{newton-start}
\\
{\bf \nabla} p^{(2)} + \rho^{(2)} {\bf \nabla} \Phi^{(0)} + \rho^{(0)}
{\bf \nabla} \Phi^{(2)} &=&
0
,
\end{eqnarray}
at $O(\varepsilon^3)$
\FL
\begin{eqnarray}
\nabla^2 \Phi^{(3)} - 4 \pi G \rho^{(3)} &=& 0
\\
{\bf \nabla} p^{(3)} + \rho^{(3)} {\bf \nabla} \Phi^{(0)} + \rho^{(0)}
{\bf \nabla} \Phi^{(3)} &=& 0
 \\
{\bf \nabla} \cdot \left[ \rho^{(0)} {\bf u}^{(3)} \right] &=& -
{\partial \rho^{(2)} \over \partial t},
\label{newton-middle}
\end{eqnarray}
and at $O(\varepsilon^4)$
\FL
\begin{eqnarray}
\label{newton-1a}
\nabla^2 \Phi^{(4)} - 4 \pi G \rho^{(4)} &=& 0 \\
{\bf \nabla} p^{(4)} + \rho^{(4)} {\bf \nabla} \Phi^{(0)} + \rho^{(0)}
{\bf \nabla} \Phi^{(4)} &=&
- \rho^{(2)} {\bf \nabla} \Phi^{(2)} \nonumber \\
&&- \rho^{(0)} {\partial {\bf u}^{(3)} \over   \partial t}  \\
{\bf \nabla} \cdot \left[ \rho^{(0)} {\bf u}^{(4)} \right] &=& -
{\partial \rho^{(3)} \over \partial t}.
\label{newton-end}
\end{eqnarray}

The general solution for the perturbation $\delta \Phi = \Phi -
\Phi^{(0)}$ to the Newtonian potential outside the star is
\beq
\delta \Phi = \sum_{J = 0}^\infty \, \sum_{m = -J}^J \, \left[ {a_{Jm}(t)
\over r^{J+1} } + b_{Jm}(t) r^J \right] Y_{Jm}(\theta,\varphi),
\label{deltaPhi}
\endeq
where $r,\theta,\varphi$ are the usual spherical polar coordinates.
We denote the eigenvalue of total angular momentum by $J$ here rather
than the more conventional $l$ in order to facilitate comparison with
the relativistic case below.  We assume that the external
gravitational field as parameterized by the coefficients $b_{Jm}$
varies with $\varepsilon$ as
\begin{eqnarray}
b_{Jm} &=& \varepsilon^2 b_{Jm}^{(2)} + \varepsilon^3 b_{Jm}^{(3)} +
\varepsilon^4 b_{Jm}^{(4)} + O(\varepsilon^5).
\label{Newtonian-b-ansatz}
\end{eqnarray}
where each $b_{Jm}^{(t)}$ varies over timescales $\propto 1 /
\varepsilon$, as in the relativistic case.  Then the coefficients
$a_{Jm}$ (the star's multipole moments) can be expanded as
\begin{eqnarray}
a_{Jm} &=& \varepsilon^2 a_{Jm}^{(2)} + \varepsilon^3 a_{Jm}^{(3)} +
\varepsilon^4 a_{Jm}^{(4)} + O(\varepsilon^5),
\label{Newtonian-a-result}
\end{eqnarray}
where the coefficients $a_{Jm}^{(t)}$ are determined from the
$b_{Jm}^{(t)}$ via Eqs.\ (\ref{newton-start}) -- (\ref{newton-end}).

There are three well-known properties of the solution (\ref{deltaPhi})
that we will
focus on.  These properties, suitably generalized, carry over to the
relativistic analysis and will play a crucial role in our matching
analysis in Sec.\ \ref{sec:response} below.
\begin{itemize}

\item
The piece of the solution with total angular momentum
eigenvalue $J$ diverges at large $r$ like $r^J$.

\item
For $J \ge 2$, the solutions are completely determined by the
coefficients $b_{Jm}^{(t)}$ at each order in perturbation theory.
{}From Eqs.\ (\ref{newton-start}) -- (\ref{newton-middle}) we have
\begin{eqnarray}
\label{Newton-fixed}
a_{Jm}^{(2)} &=& k_J R^{2J + 1} b_{Jm}^{(2)} \\
a_{Jm}^{(3)} &=& k_J R^{2J + 1} b_{Jm}^{(3)},
\label{Newton-fixed-1}
\end{eqnarray}
where $R$ is the stellar radius and $k_J$ is a fixed dimensionless
constant which is determined by the stellar equation of state.
At higher orders in $\varepsilon$ things are a little more
complicated.  From the form of Eqs.\ (\ref{newton-1a}) -- (\ref{newton-end})
it follows
that $a_{Jm}^{(4)}$ consists of a piece linear in
${\ddot b}_{Jm}^{(2)}$ and $b_{Jm}^{(4)}$,
together with a term
\beq
\sum_{J'\,m'\,J''\,m''}\,
K_{J'\,m'\,J''\,m''}^{Jm} \, b_{J'm'}^{(2)} \, b_{J''m''}^{(2)},
\label{Newtonian-nonlinear}
\endeq
for some constants $K_{J'\,m'\,J''\,m''}^{Jm}$.
Similarly the
solutions inside the star are completely determined by the
coefficients $b_{Jm}^{(t)}$.

\item
The $J= 0,1$ pieces of the solution outside the star contain physical
information only
at $O(\varepsilon^4)$.  Newtonian conservation of mass forbids
solutions with $a_{00} \ne 0$, and $b_{00}$ is an additive constant to
the potential which has no physical consequences.  By a change of in
the origin of coordinates of the from ${\bf r} \to {\bf r} - {\bf
r}_0$ which makes ${\bf r}=0$ the center of mass of the star, one can
enforce $a_{1m}^{(t)} = 0$ for $t = 2,3,4$.  This then implies that
$b_{1m}^{(2)} = b_{1m}^{(3)} = 0$.  However, the
coefficient $b_{1m}^{(4)}$ can be nonzero; this coefficient encodes the
acceleration of the center of mass of the star with respect to
inertial reference frames (due to couplings of
multipole moments of the star to multipole moments of the external
gravitational field with $J \ge 2$).
Similarly, inside the star, $J=0$ perturbations and $J=1$
perturbations to a rotating state are driven by interactions with the
external gravitational field only at order $O(\varepsilon^4)$.

\end{itemize}

\subsection{The relativistic case}
\label{internal-scheme:relativistic}

Turn now to the corresponding analysis in the relativistic case.
Let ${\cal G}$ denote the space of solutions of the equations of
motion (\ref{fluid2}) -- (\ref{hB4}), consisting of the tensor fields
${\hat h}_{ab}^{(t)}$ for $2 \le t \le 4$ together with the fluid
variables.  The gauge transformations (\ref{gauge2}) -- (\ref{gauge4})
define an equivalence relation on ${\cal G}$.  Let ${\cal S}$ be the
set of equivalence classes, which is the physical solution space.

We start by showing that there is a well defined action of the
rotation group on ${\cal S}$.  Let $\psi_R : M^{({\rm NS})} \to M^{({\rm
NS})}$ be the isometry of the background spacetime corresponding to a
rotation $R$.  Under this rotation an element $({\hat h}_{ab}^{(2)},
h_{ab}^{(3)}, {\hat h}_{ab}^{(4)}, \ldots)$ of ${\cal G}$ gets mapped
to $(\psi_{R\,*} {\hat h}_{ab}^{(2)},
\psi_{R\,*} h_{ab}^{(3)}, \psi_{R\,*} {\hat h}_{ab}^{(4)}, \ldots)$.
{}From Eqs.\ (\ref{gauge2}) -- (\ref{gauge4}) and using $\psi_{R\,*}
g_{ab}^{({\rm NS})} = g_{ab}^{({\rm NS})}$ it can be seen that if two
elements of ${\cal G}$ are related by a gauge transformation generated
by the vector fields $\xi^{(2)\,a}, \xi^{(3)\,a}$, and $\xi^{(4)\,a}$, then
their images under $R$ are also gauge equivalent, related by the gauge
transformation generated by $\psi_{R\,*} \xi^{(2)\,a}, \psi_{R\,*}
\xi^{(3)\,a}$, and $\psi_{R\,*} \xi^{(4)\,a}$.  Thus the action of the
rotation group on ${\cal G}$ extends to an action on ${\cal S}$.  If
$J_i$ is the generator of the action of rotations on ${\cal S}$, then
we can classify elements of ${\cal S}$ in the usual way according to
the eigenvalue $J(J+1)$ of $J_i J^i$ and the eigenvalue $m$ of $J_z$.

Next, we define the electric and magnetic pieces of the Riemann tensor
in the usual way:
\beq
{\cal E}_{ab} = R_{acbd} {\hat t}^c {\hat t}^d
\endeq
and
\beq
{\cal B}_{ab} = {1 \over 2} \epsilon_{cade} R^{de}_{\ \
bf} {\hat t}^c {\hat t}^f.
\endeq
Here $t^a = t^a / ||t^a||$ and all quantities are defined with
respect to the metric $g_{ab}^{({\rm INTERIOR})}$ of Eq.\
(\ref{interiormetric1}).  In our chosen gauge (\ref{gaugechoice}),
${\cal E}_{ab}$ and ${\cal B}_{ab}$ will be purely
spatial in the coordinate system (\ref{neutronstar1}), so we write
these tensors as ${\cal E}_{ij}$ and ${\cal B}_{ij}$.  We will use these
variables to discuss the large $r$ behavior of the solutions because
they are more nearly gauge invariant than the metric perturbations.
As usual we expand these tensors as \cite{noteEB}
\beq
{\cal E}_{ij} = {\cal E}_{ij}^{(0)} + \varepsilon^2 {\cal
E}_{ij}^{(2)} + \varepsilon^3 {\cal E}_{ij}^{(3)} + \varepsilon^4
{\cal E}_{ij}^{(4)} + O(\varepsilon^5),
\endeq
and
\beq
{\cal B}_{ij} = \varepsilon^2 {\cal
B}_{ij}^{(2)} + \varepsilon^3 {\cal B}_{ij}^{(3)} + \varepsilon^4
{\cal B}_{ij}^{(4)} + O(\varepsilon^5).
\endeq
These tensors transform via the gauge transformation rule
(\ref{gauge2}) -- (\ref{gauge4}) under the spatial gauge
transformations allowed by Eq.\ (\ref{spatial-gauge}).  The tensors
${\cal B}_{ij}^{(2)}$ and ${\cal B}_{ij}^{(3)}$ are gauge invariant,
but the rest are gauge dependent.

Suppose now that one is given an element of ${\cal S}$ corresponding to
the eigenvalues $(J,m)$.  Then it is straightforward to show that one
can choose a gauge in which ${\cal E}_{ij}$ and ${\cal B}_{ij}$
are eigenfunctions of ${\bar J}_i {\bar J}^i$ and ${\bar J}_z$
with the corresponding eigenvalues, where ${\bar J}_i$ is the
generator of the action of rotations on ${\cal G}$.  Below, when
discussing the $(J,m)$ sector of ${\cal S}$, we will always assume
that such a gauge has been chosen.

To discuss the general form of the solutions it is convenient to
use the pure orbital tensor spherical harmonics $T_{ij}^{\lambda
L,Jm}(\theta,\varphi)$ of Thorne, which are defined in Eqs.\
(2.40a) -- (2.40f) of Ref.\ \cite{ThorneRMP}.  Here $\lambda$,
$L$, $J$ and $m$ are integers with $\lambda = 0$ or $2$, $0
\le J < \infty$, $-J \le m \le J$, and \cite{noteRMP}
\begin{equation}
\left. \begin{array}{ll} J-2 \le L \le J+2 & {\rm \ \ for} \ \ J \ge 2
\nonumber \\
		{\ \ \ \ \ \,}1 \le L \le 3 & {\rm \ \ for} \ \ J =1
\\
\label{lprimerange}
		{\ \ \ \ \ \ \ \ }L=2 &  {\rm \ \ for} \ \ J =0 \nonumber \\
        \end{array} \right\}. \,\,\,
\end{equation}
These tensors are complete in the sense that any symmetric tensorial
function $f_{ij}(\theta,\varphi)$ on the unit sphere can be expanded as
\beq
f_{ij}(\theta,\varphi) = \sum_{\lambda=0,2} \ \sum_{J=0}^\infty \sum_{m
= -J}^J \sum_{L} \, f_{\lambda L,Jm} \, T^{\lambda
L,Jm}_{ij}(\theta,\varphi).
\endeq
As the notation suggests, the tensor $T_{ij}^{\lambda
L,Jm}(\theta,\varphi)$ is an eigenfunction of ${\bar J}_i {\bar J}^i$
with eigenvalue $J(J+1)$ and of ${\bar J}_z$ with eigenvalue $m$.  It
is also an
eigenfunction of ${\bar L}_i {\bar L}^i$ with eigenvalue $L(L+1)$,
where ${\bar L}_i$ is the generator of the ``pure orbital''
representation of the rotation group on symmetric 2-index tensors
\cite{ThorneRMP}.
Using these tensors we can write down the general form for ${\cal
E}_{ij}^{(t)}$ and ${\cal B}_{ij}^{(t)}$ on the $(J,m)$ sector of the
solution space ${\cal S}$:
\beq
{\cal E}_{ij}^{(t)} =
\sum_{\lambda=0,2} \ \sum_L {\cal E}_{\lambda L,Jm}^{(t)} \, T^{\lambda
L,Jm}_{ij}(\theta,\varphi)
\label{calE-general}
\endeq
and
\beq
{\cal B}_{ij}^{(t)} = \sum_{\lambda=0,2} \ \sum_L {\cal B}_{\lambda
L,Jm}^{(t)} \, T^{\lambda L,Jm}_{ij}(\theta,\varphi).
\label{calB-general}
\endeq
Here the coefficients are functions of $t$ and $r$, where
$(t,r,\theta,\varphi)$ is the coordinate system (\ref{neutronstar1}), and
the summation over $L$ is over the range (\ref{lprimerange}).
Note that the eigenvalue $L$ is well defined on ${\cal G}$ but not on the
physical solution space ${\cal S}$, since the ``pure orbital''
representation of the rotation group does not extend from ${\cal G}$
to ${\cal S}$.

\subsubsection{Leading order solutions}
\label{leading-order}

Consider now the general form of the linearized solutions ${\cal
E}_{ij}^{(2)}$ and ${\cal B}_{ij}^{(2)}$ outside the star.  From Eq.\
(\ref{hB2}) these describe a stationary perturbation of the
Schwarschild spacetime \cite{stationary}.  In the flat spacetime limit
$M \to 0$, ${\cal E}_{ij}^{(2)}$ obeys the system of equations
\cite{Maartens}
\begin{equation}
{\cal E}^{(2)}_{[ij]} = {\cal E}^{(2)\,i}_i  = D^i {\cal E}^{(2)}_{ij} =
D_{[i} {\cal E}^{(2)}_{j]k} = 0.
\endeq
Using the methods of Ref.\ \cite{ThorneRMP} one can
show that the general solution for ${\cal E}^{(2)}_{ij}$ is
\begin{eqnarray}
{\cal E}^{(2)}_{ij} &=& \sum_{J=0}^\infty \sum_{m = -J}^J \ {a_{Jm}^{(2)}
\over r^{J + 3}} \, \, T_{ij}^{2\,J+2,Jm} \nonumber \\
&& + \sum_{J=2}^\infty \sum_{m = -J}^J \ b_{Jm}^{(2)} r^{J -2} \,
\, T_{ij}^{2\,J-2,Jm},
\label{calE-flat}
\end{eqnarray}
which is none other than $\partial_i \partial_j \, \Phi^{(2)}$,
where $\Phi^{(2)}$ is the $O(\varepsilon^2)$ piece of the Newtonian
potential (\ref{deltaPhi}).  Note that the summation in the second
term in Eq.\ (\ref{calE-flat}) starts at $J=2$, unlike the Newtonian
formula.  Also the term proportional to $a_{1m}^{(2)}$ is pure gauge.
Similarly ${\cal B}_{ij}^{(2)}$ satisfies
\cite{Maartens}
\begin{equation}
{\cal B}^{(2)}_{[ij]} = {\cal B}^{(2)\,i}_i  = D^i {\cal B}^{(2)}_{ij} =
D_{[i} {\cal B}^{(2)}_{j]k} = 0,
\label{calB-flat-eqns}
\endeq
and can be written as
\begin{eqnarray}
{\cal B}^{(2)}_{ij} &=& \sum_{J=1}^\infty \sum_{m = -J}^J \ {c_{Jm}^{(2)}
\over r^{J + 3}} \, \, T_{ij}^{2\,J+2,Jm} \nonumber \\
&& + \sum_{J=2}^\infty \sum_{m = -J}^J \ d_{Jm}^{(2)} r^{J -2} \,
\, T_{ij}^{2\,J-2,Jm}.
\label{calB-flat}
\end{eqnarray}
In Eq.\ (\ref{calB-flat}) there is no $J=0$ term in the
first sum; such a term would satisfy Eqs.\ (\ref{calB-flat-eqns})
but is not present for metric perturbations that
satisfy the original linearized Einstein equation.

Returning from the flat spacetime limit, the general solutions for
${\cal E}_{ij}^{(2)}$ and ${\cal B}_{ij}^{2}$ in the case $M \ne 0$
outside the star can again be parameterized by coefficients
$a_{Jm}^{(2)}$, $b_{Jm}^{(2)}$, $c_{Jm}^{(2)}$, and $d_{Jm}^{(2)}$:
\begin{eqnarray}
{\cal E}^{(2)}_{ij} &=& \sum_{\begin{array}{ll}J&=0\\
J&\ne 1
\end{array}
}^\infty \sum_{m = -J}^J \
a_{Jm}^{(2)}\,
{}^{(\uparrow)}{\cal E}_{ij}^{Jm}
\nonumber \\
&&
+ \sum_{J=2}^\infty \sum_{m = -J}^J \ b_{Jm}^{(2)} \,
{}^{(\downarrow)}{\cal E}_{ij}^{Jm},
\label{calE-curved}
\end{eqnarray}
and
\begin{eqnarray}
{\cal B}^{(2)}_{ij} &=& \sum_{J=1}^\infty \sum_{m = -J}^J \
c_{Jm}^{(2)}\,
{}^{(\uparrow)}{\cal B}_{ij}^{Jm}
\nonumber \\
&&
+ \sum_{J=2}^\infty \sum_{m = -J}^J \ d_{Jm}^{(2)} \,
{}^{(\downarrow)}{\cal B}_{ij}^{Jm}.
\label{calB-curved}
\end{eqnarray}
Here ${}^{(\uparrow)}{\cal E}_{ij}^{Jm}$, ${}^{(\downarrow)}{\cal
E}_{ij}^{Jm}$, ${}^{(\uparrow)}{\cal B}_{ij}^{Jm}$, and
${}^{(\downarrow)}{\cal B}_{ij}^{Jm}$ are certain fixed linearly
independent solutions of the perturbation equations in the $(J,m)$
sector whose large $r$ behavior is given by \cite{question}
\begin{eqnarray}
{}^{(\uparrow)}{\cal E}_{ij}^{Jm} &=&
{1 \over r^{J + 3}} \, \, T_{ij}^{2\,J+2,Jm} + O\left({M \over r^{J+4}}\right),
\label{calE-up}
 \\ \mbox{}
{}^{(\downarrow)}{\cal E}_{ij}^{Jm} &=& r^{J -2} \, T_{ij}^{2\,J-2,Jm}
+ O( M r^{J-3} ),
\label{calE-down}
\\ \mbox{}
{}^{(\uparrow)}{\cal B}_{ij}^{Jm} &=&
{1 \over r^{J + 3}} \, \, T_{ij}^{2\,J+2,Jm} + O\left({M \over
r^{J+4}}\right),
\label{calB-up}
\\ \mbox{}
{}^{(\downarrow)}{\cal B}_{ij}^{Jm} &=& r^{J -2} \, T_{ij}^{2\,J-2,Jm}
+ O( M r^{J-3} ).
\label{calB-down}
\end{eqnarray}
These quantities are functions of $r,\theta$ and $\varphi$ only, and can
be expressed as expansions of the form (\ref{calE-general}) and
(\ref{calB-general}).  The functions ${}^{(\uparrow)}{\cal
B}_{ij}^{Jm}$ and ${}^{(\downarrow)}{\cal B}_{ij}^{Jm}$ are gauge
independent, while ${}^{(\uparrow)}{\cal E}_{ij}^{Jm}$ and
${}^{(\downarrow)}{\cal E}_{ij}^{Jm}$ are gauge-dependent, their large
$r$ forms (\ref{calE-up}) and (\ref{calE-down}) being achieved only in
certain gauges \cite{notegauge}.  In Eq.\ (\ref{calE-curved}) we have
omitted the $J=1$ term from the first summation since it is pure gauge.

The above analysis assumes that the $(J,m)$ sector of the space of
stationary linear perturbations off Schwarschild, modulo gauge
transformations, is of dimension 4 for $J \ge 2$, of dimension $1$ for
$J=1$, and of dimension 1 for $J=0$ \cite{stationary}.   These
dimensionalities are strongly suggested by the form (\ref{calE-flat}) and
(\ref{calB-flat}) of the solutions in the $M \to 0$ limit.  To
rigorously prove these dimensionalities in the $J \ge 2$ case, one can
appeal to the Newman-Penrose perturbation formalism.  The
gauge-invariant Newman-Penrose coefficient $\Psi_4$ determines the
perturbation
uniquely up to gauge transformations and up to $J=0,1$ perturbations
\cite{Waldproof,Chzranowski}.  In the zero frequency case of interest
to us, $\Psi_4$ can be written as
\beq
\Psi_4(r,\theta,\varphi) = {1 \over r^4} \sum_{Jm} R_{Jm}(r)
\ {}_{-2}Y_{Jm}(\theta,\varphi),
\endeq
where $R_{Jm}(r)$ obeys the zero-frequency Teukolsky equation
\cite{Teukolsky}.  The Teukolsky equation has two complex (or four real)
linearly independent solutions, since it is a second order ordinary
differential equation \cite{noteT}, which proves the result.  The
dimensionalities in the cases $J=0,1$ can be proved directly; see,
e.g., Ref.\ \cite{Burko}.

The three properties of the Newtonian solutions discussed in Sec.\
\ref{internal-scheme:Newtonian} have
analogs in their relativistic counterparts (\ref{calE-curved}) and
(\ref{calB-curved}).  First, in the sector $(J,m)$,
the Riemann tensor perturbation diverges at large $r$ like $r^{J-2}$,
corresponding to a rate of divergence of the metric perturbation (in a
suitably chosen gauge) of $r^J$.
Second, for $J \ge 2$, the solutions are completely determined (up to
gauge) by the
coefficients $b_{Jm}^{(2)}$ and $d_{Jm}^{(2)}$ which parameterize the
external gravitational perturbations.
The perturbed fluid
equations (\ref{fluid2}) and (\ref{hB2})
determine the response of the star and the coefficients $a_{Jm}^{(2)}$
and $c_{Jm}^{(2)}$
as functions of $b_{Jm}^{(2)}$ and $d_{Jm}^{(2)}$:
\begin{eqnarray}
a_{Jm}^{(2)}(t) &=& k^\prime_J M^{2 J + 1} b_{Jm}^{(2)}(t)
\label{el1}
\\
c_{Jm}^{(2)}(t) &=& k^{\prime\prime}_J M^{2 J + 1} d_{Jm}^{(2)}(t).
\label{mg1}
\end{eqnarray}
This is completely analogous to the Newtonian case
(\ref{Newton-fixed}) except that there
are now the magnetic type moments (\ref{mg1}) in addition to the
electric type moments (\ref{el1}).
Here $k^\prime_J$ and $k^{\prime\prime}_J$ are dimensionless constants
determined by the stellar equation of state.  Similarly the fluid
variables $T_{ab}^{(2)}$ are determined.

Finally, consider the status of the $J
= 0,1$ sectors.
Solutions of the
perturbation equations (\ref{fluid2}) and (\ref{hB2}) which are purely
$J=0$ do exist.  If we denote the background solution with total mass
$M$ as $g_{ab}^{({\rm NS})}[M]$ and $T_{ab}[M]$, then
the quantities $g_{ab}^{({\rm NS})}[M + \varepsilon^2 \delta M(t)]$
and $T_{ab}[M + \varepsilon^2 \delta M(t)]$ satisfy the perturbation
equations to $O(\varepsilon^2)$ for any choice of $\delta M(t)$.
However, in this case no solutions to the higher order perturbation equations
(\ref{fluid3}), (\ref{fluid4}) and (\ref{hB3}) and (\ref{hB4}) exist
unless $\delta M(t) =0$.  Thus there are no $J=0$ perturbations at
order $O(\varepsilon^2)$.  Alternatively, one can argue that
$J=0$ perturbations are forbidden at this order in
$\varepsilon$ by conservation of baryon number \cite{note9}.

A similar situation holds in the $J=1$ sector.  The
$J=1$ piece of the solutions (\ref{calE-curved}) and
(\ref{calB-curved}) is parameterized by the coefficient
$c_{1m}^{(2)}$, since the $a_{1m}^{(2)}$ term is pure gauge and
there are no $J=1$ terms in the pieces of the solutions that diverge
at large $r$.  This coefficient $c_{1m}^{(2)}$ parameterizes a
perturbation to a rotating state.  The second order perturbation
equations are satisfied by any choice of time dependent coefficient
$c_{1m}^{(2)}(t)$, but, as in the $J=0$ case, the higher order
perturbation equations can only be satisfied if $c_{1m}^{(2)}(t) = 0$
\cite{proof1}.

To summarize, if we now combine Eqs.\ (\ref{calE-curved}),
(\ref{calB-curved}), (\ref{el1}) and (\ref{mg1}) and
revert to using metric perturbations instead of curvature
perturbations, we can write the
solutions for ${\hat h}_{ab}^{(2)}$ and $T_{ab}^{(2)}$ as
\beq
{\hat h}_{ab}^{(2)} = \sum_{J=2}^\infty \sum_m \left[ b_{Jm}^{(2)} {\hat
h}^{E,Jm}_{ab} + d_{Jm}^{(2)} {\hat h}_{ab}^{B,Jm} \right]
\label{linear-solns-h}
\endeq
and
\beq
T_{ab}^{(2)} = \sum_{J=2}^\infty \sum_m \left[ b_{Jm}^{(2)} T^{E,Jm}_{ab} +
d_{Jm}^{(2)} T_{ab}^{B,Jm} \right].
\label{linear-solns-T}
\endeq
Here the electric-type quantities ${\hat h}_{ab}^{E,Jm}$ and $T_{ab}^{E,Jm}$
and magnetic-type quantities ${\hat h}_{ab}^{B,Jm}$ and $T_{ab}^{B,Jm}$ are
fixed up to gauge transformations.

\subsubsection{Higher order solutions}

Consider next the $O(\varepsilon^3)$ piece of the solution space
${\cal S}$.  The equations (\ref{fluid3}) and (\ref{hB3}) satisfied by
the metric perturbation ${\hat h}_{ab}^{(3)}$ and by the
$O(\varepsilon^3)$ fluid variables are of the same form as the
$O(\varepsilon^2)$ equations, except that they contain source terms
which are time derivatives of the $O(\varepsilon^2)$ perturbations.
Since ${\hat h}_{ab}^{(3)}$ is therefore not a vacuum perturbation of
Schwarschild, one cannot use the argument of Sec.\ \ref{leading-order}
to determine the general solution.  However, we can instead make the
following argument.

Fix a specific $O(\varepsilon^3)$ solution $({\hat h}_{ab}^{(2)}, {}^0
{\hat h}_{ab}^{(3)}, T_{ab}^{(2)}, {}^0 T_{ab}^{(3)})$.  Then, any
other solution $({\hat h}_{ab}^{(2)}, {\hat h}_{ab}^{(3)},
T_{ab}^{(2)}, T_{ab}^{(3)})$ which has the same
$O(\varepsilon^2)$ part must be of the form
\begin{eqnarray}
{\hat h}_{ab}^{(3)} &=& {}^0 {\hat h}_{ab}^{(3)} + \Delta {\hat
h}_{ab}^{(3)},
\label{hath3-general}
\\
\mbox{}
T_{ab}^{(3)} &=& {}^0 T_{ab}^{(3)} + \Delta T_{ab}^{(3)},
\label{T3-general}
\end{eqnarray}
where from Eqs.\ (\ref{fluid3}) and (\ref{hB3}) the differences
$\Delta {\hat h}_{ab}^{(3)}$ and
$\Delta T_{ab}^{(3)}$ obey the same equations (\ref{fluid2}) and
(\ref{hB2}) as the $O(\varepsilon^2)$ variables.
Hence from Eqs.\ (\ref{linear-solns-h}) and (\ref{linear-solns-T}) we
can write
\begin{eqnarray}
{\hat h}_{ab}^{(3)} &=& {}^0 {\hat h}_{ab}^{(3)} +
\sum_{J=2}^\infty \sum_m \left[ b_{Jm}^{(3)} {\hat h}^{E,Jm}_{ab} +
d_{Jm}^{(3)} {\hat h}_{ab}^{B,Jm} \right]
\label{hath3-general-1}
\\
\mbox{}
T_{ab}^{(3)} &=& {}^0 T_{ab}^{(3)} + \sum_{J=2}^\infty \sum_m \left[
b_{Jm}^{(3)} T^{E,Jm}_{ab} + d_{Jm}^{(3)} T_{ab}^{B,Jm} \right],
\label{T3-general-1}
\end{eqnarray}
for some coefficients $b_{Jm}^{(3)}$, $d_{Jm}^{(3)}$.
The quantities ${}^0 {\hat
h}_{ab}^{(3)}$ and ${}^0 T_{ab}^{(3)}$ here
can be regarded as fixed linear functions of the
time derivatives of ${\hat h}_{ab}^{(2)}$ and $T_{ab}^{(2)}$, or
equivalently as fixed linear functions of the time derivatives ${\dot
b}_{Jm}^{(2)}$ and ${\dot d}_{Jm}^{(2)}$.  Thus, the
$O(\varepsilon^3)$ solutions are completely specified by giving the
coefficients $b_{Jm}^{(2)}$, $d_{Jm}^{(2)}$, $b_{Jm}^{(3)}$ and
$d_{Jm}^{(3)}$ as functions of time.

A similar argument can be used at $O(\varepsilon^4)$.  Here the
equations of motion (\ref{fluid4}) and (\ref{hB4}) for the variables
${\hat h}_{ab}^{(4)}$ and $T_{ab}^{(4)}$ contain source terms which
are linear in time derivatives of the lower order variables, and also
source terms which are quadratic in the $O(\varepsilon^2)$ variables.
The argument shows that
\begin{eqnarray}
{\hat h}_{ab}^{(4)} &=& {}^0 {\hat h}_{ab}^{(4)} +
\sum_{J=2}^\infty \sum_m \left[ b_{Jm}^{(4)} {\hat h}^{E,Jm}_{ab} +
d_{Jm}^{(4)} {\hat h}_{ab}^{B,Jm} \right]
\label{hath4-general-1}
\\
\mbox{}
T_{ab}^{(4)} &=& {}^0 T_{ab}^{(4)} + \sum_{J=2}^\infty \sum_m \left[
b_{Jm}^{(4)} T^{E,Jm}_{ab} + d_{Jm}^{(4)} T_{ab}^{B,Jm} \right],
\label{T4-general-1}
\end{eqnarray}
for some coefficients $b_{Jm}^{(4)}$, $d_{Jm}^{(4)}$.
Here the quantities
${}^0 {\hat h}_{ab}^{(4)}$ and ${}^0 T_{ab}^{(4)}$ depend linearly on
${\ddot b}_{Jm}^{(2)}$, ${\ddot d}_{Jm}^{(2)}$, ${\dot b}_{Jm}^{(3)}$,
${\dot d}_{Jm}^{(3)}$, and quadratically on $b_{Jm}^{(2)}$ and
$d_{Jm}^{(2)}$ as in Eq.\ (\ref{Newtonian-nonlinear}).

Note that in Eqs.\ (\ref{hath3-general-1}) -- (\ref{T4-general-1}) we
have not specified yet how the fixed solutions ${}^{0} {\hat
h}_{ab}^{(3)}$, ${}^0 T_{ab}^{(3)}$, ${}^{0} {\hat h}_{ab}^{(4)}$, and
${}^0 T_{ab}^{(4)}$ are chosen.  One has the freedom to add to these
functions any linearized solutions of the form (\ref{linear-solns-h}) and
(\ref{linear-solns-T}).  This has the effect of redefining the
zero points of the variables $b_{Jm}^{(3)}$, $d_{Jm}^{(3)}$,
$b_{Jm}^{(4)}$, and $d_{Jm}^{(4)}$.  It will turn out below that for
the calculations in this paper we shall not need to resolve this
ambiguity.

\section{RESPONSE OF THE STAR TO EXTERNAL FIELDS}
\label{sec:response}

In this section we will perform the matching calculations outlined in
Sec.\ \ref{matchingscheme} above to determine the response of the
neutron star to the external perturbing gravitational fields.

\subsection{Change in central density}
\label{response:centraldensity}

The change in the structure of the neutron star is described by
the stress tensor perturbations $T_{ab}^{(2)}$, $T_{ab}^{(3)}$ and
$T_{ab}^{(4)}$.  Following Refs.~\cite{BradyHughes,Wiseman}, we focus
in particular on the {\it central density} $\rho_c$ of the star, since
changes in the central density reflect star-crushing forces
that tend to destabilize the star to radial collapse, or
``anti--crushing'' forces that tend to stabilize the star against
radial collapse.  We shall show in this subsection that the
change in central density depend only on the details of the matchings
at the orders $0 \le t \le 2$, and is independent of the details of
the matchings for $t \ge 3$ (see Fig.\ \ref{fig:matching} below).
This fact will considerably simplify our analysis.

Following Eq.~(\ref{stresstensor1}) we expand the density $\rho$ as
\begin{equation}
\rho = \rho^{(0)} + \varepsilon^2 \rho^{(2)} + \varepsilon^3
\rho^{(3)} + \varepsilon^4 \rho^{(4)} + O(\varepsilon^5).
\end{equation}
We define the central density $\rho_c(t)$ to be the maximum value of
the density in the star on a hypersurface of constant
$t(\varepsilon)$.  In our chosen gauge (\ref{gaugechoice}), this is the same
as a hypersurface of constant $t$, where $t$ is the time coordinate in
Eq.\ (\ref{neutronstar1}).  Note that $\rho_c(t)$ is not the same as
$\rho(t,r=0)$, since the location of the star's center can be changed by the
perturbation \cite{BradyHughes}.  Since $\rho^{(0)}$ has a local
maximum at $r=0$, we find that the change in central density is given by
\begin{eqnarray}
\delta \rho_c(t) &=& \varepsilon^2 \rho^{(2)}(t,0) +
\varepsilon^3 \rho^{(3)}(t,0) \nonumber \\
\mbox{} && + \varepsilon^4 \left[ \rho^{(4)}(t,0) - {3 \over 2} { D_i
\rho^{(2)}(t,0) D^i \rho^{(2)}(t,0) \over D^2 \rho^{(0)}(t,0)} \right]
\nonumber \\
\mbox{}  && + O(\varepsilon^5).
\label{rhocbasic}
\end{eqnarray}
Here $D_i$ is the spatial derivative associated with the background
metric $g_{ab}^{({\rm NS})}$.  We will now show that the first two
terms in the expansion (\ref{rhocbasic}) are vanishing.

To prove this result, we modify slightly an argument used
by Brady and Hughes in a similar context \cite{BradyHughes}.  The
change in the stellar structure can be decomposed into contributions from
each of the $(J,m)$ sectors of the perturbation [cf.\ Sec.\
\ref{internal-scheme:relativistic} above].  Therefore we can express
the density
perturbations $\rho^{(t)}$ for $2 \le t \le 4$ as
\begin{equation}
\rho^{(t)}(t,r,\theta,\varphi) = \sum_{Jm} \rho^{(t)}_{Jm}(t,r)
Y_{Jm}(\theta,\varphi),
\end{equation}
where $(t,r,\theta,\varphi)$ are Schwarschild-like coordinates adapted
to the unperturbed, spherical star.  Since each of the density
perturbations $\rho^{(t)}$
is a smooth function of position at $r=0$, it
follows that $\rho^{(t)}_{Jm}(t,0) =0$ for all $J,m$ except possibly
for the spherically symmetric, $J=m=0$ sector, as shown by Brady and
Hughes \cite{BradyHughes}.  Thus, only spherically symmetric sector can
change the central density.

Next, from the analysis of Sec.\ \ref{internal-scheme:relativistic}
above it follows
that the $O(\varepsilon^2)$ perturbation contains only $J \ge 2$
excitations, and
has no $J=0$ parts [see Eqs.\ (\ref{linear-solns-h}) and
(\ref{linear-solns-T})].  It follows that $\rho^{(2)}(t,0) =0$.
Similarly, at $O(\varepsilon^3)$, the second terms in Eqs.\
(\ref{hath3-general-1}) and (\ref{T3-general-1}) give a vanishing
contribution to $\rho^{(3)}(t,0)$.  The terms ${}^0 {\hat
h}^{(3)}_{Jm}$ and ${}^{0} T_{ab}^{(3)}$ in Eqs.\
(\ref{hath3-general-1}) and (\ref{T3-general-1}) are linear functions
of the time derivatives of the $O(\varepsilon^2)$ variables ${\hat
h}_{ab}^{(2)}$ and $T_{ab}^{(2)}$; since those variables have no $J=0$
parts, neither do ${}^0 {\hat h}^{(3)}_{Jm}$ and ${}^{0} T_{ab}^{(3)}$.
Therefore there is no $J=0$ part to the $O(\varepsilon^3)$
perturbation and hence $\rho^{(3)}(t,0) =0$.  In addition, the
quantity $D_i \rho^{(2)}(t,0)$ appearing in Eq.\ (\ref{rhocbasic})
must vanish, as there is no $J=1$ piece to the $O(\varepsilon^2)$
perturbation.  Hence, we can rewrite Eq.\ (\ref{rhocbasic}) as
\begin{eqnarray}
\delta \rho_c(t) &=&
\varepsilon^4 \rho^{(4)}(t,0) + O(\varepsilon^5).
\label{rhocbasic1}
\end{eqnarray}

Finally, consider the decompositions (\ref{hath4-general-1}) and
(\ref{T4-general-1}) of the fourth order perturbation variables.  The
$t=4$ matchings can affect only the second terms in these
equations, and those terms will not contribute to the $J=0$ part of
the perturbation since they have only $J \ge 2$ parts.
In addition, the only $J=0$ piece of the terms ${}^0 {\hat
h}_{ab}^{(4)}$ and ${}^0 T_{ab}^{(4)}$ are due to the source
terms in Eqs.\ (\ref{forcedef1}) and (\ref{hB4}) which are quadratic
in the $O(\varepsilon^2)$ variables, since the linear terms will not
have any $J=0$ part by the argument of the last paragraph.  Thus, the
change in central density $\rho^{(4)}(t,0)$ is a quadratic function of
the $O(\varepsilon^2)$ perturbation variables and is independent of
the details of the $t=3$ and $t=4$ matching.

\subsection{Matching calculations}
\label{response:matching}

Turn now to the matching of the internal and external solutions, which
was outlined in Sec.\ \ref{matchingscheme} above.  We need to match the
metrics $g_{ab}^{({\rm INTERIOR})}$ with $\varphi_* g_{ab}^{({\rm
EXTERIOR})}$ order by order in a double expansion in the parameters
$M/r$ and $r/{\cal R}$.  The matching scheme is illustrated in Fig.\
\ref{fig:matching}.

The matchings for the cases $(s,t) = (0,0), (0,1),$ $ (0,2), (1,0),
(1,1), (1,2)$ have been explicitly worked out by Mino et.\ al.\
\cite{Mino}, in the case where the interior metric describes a black
hole \cite{noteJ0}.  Their results are also applicable to the neutron
star case.
They show that ${\hat h}_{ab}^{(1)}$ vanishes identically, and obtain
the constraints on the embedding free
functions that ${}^{(0,0)} {\cal F}^\alpha(t)$ describes a geodesic of
$g_{ab}^{(B)}$, and that
\begin{equation}
{}^{(1,0)} {\cal F}_i^0(t) = 0, \ \ \ \ \ \ \ \
{}^{(1,0)} {\cal F}_i^k(t) = \delta_i^k.
\label{embeddingf1}
\end{equation}

Consider now the determination of the metric perturbation ${\hat
h}_{ab}^{(2)}$.  From Sec.\ \ref{internal-scheme:relativistic} we know that
${\hat h}_{ab}^{(2)}$ contains only $J\ge2$ pieces, and that any
$(J,m)$ piece of ${\hat h}_{ab}^{(2)}$ must diverge at large $r$ like
$r^J$.  However, from Fig.\ \ref{fig:matching} it can be seen that
${\hat h}_{ab}^{(2)}$ diverges at large $r$ like $r^2$.  Hence, ${\hat
h}_{ab}^{(2)}$ must be purely $J=2$.  There are thus 10 free coefficients
$b_{2m}^{(2)}$ and $d_{2m}^{(2)}$ in Eqs.\ (\ref{linear-solns-h}) and
(\ref{linear-solns-T}) which determine ${\hat h}_{ab}^{(2)}$.  These
parameters are determined by the $(s,t) = (0,2)$ matching in Fig.\
\ref{fig:matching}, which using Eq.\ (\ref{embeddingf1}) simply
dictates that the constant asymptotic
values of the curvatures ${\cal E}^{(2)}_{ij}$ and ${\cal
B}^{(2)}_{ij}$ associated with ${\hat h}_{ab}^{(2)}$ agree with
the ${\cal E}_{ij}$ and ${\cal B}_{ij}$ of the background metric
$g_{ab}^{(B)}$ evaluated on the geodesic.

{\vskip 1cm}
\begin{figure}
{\psfig{file=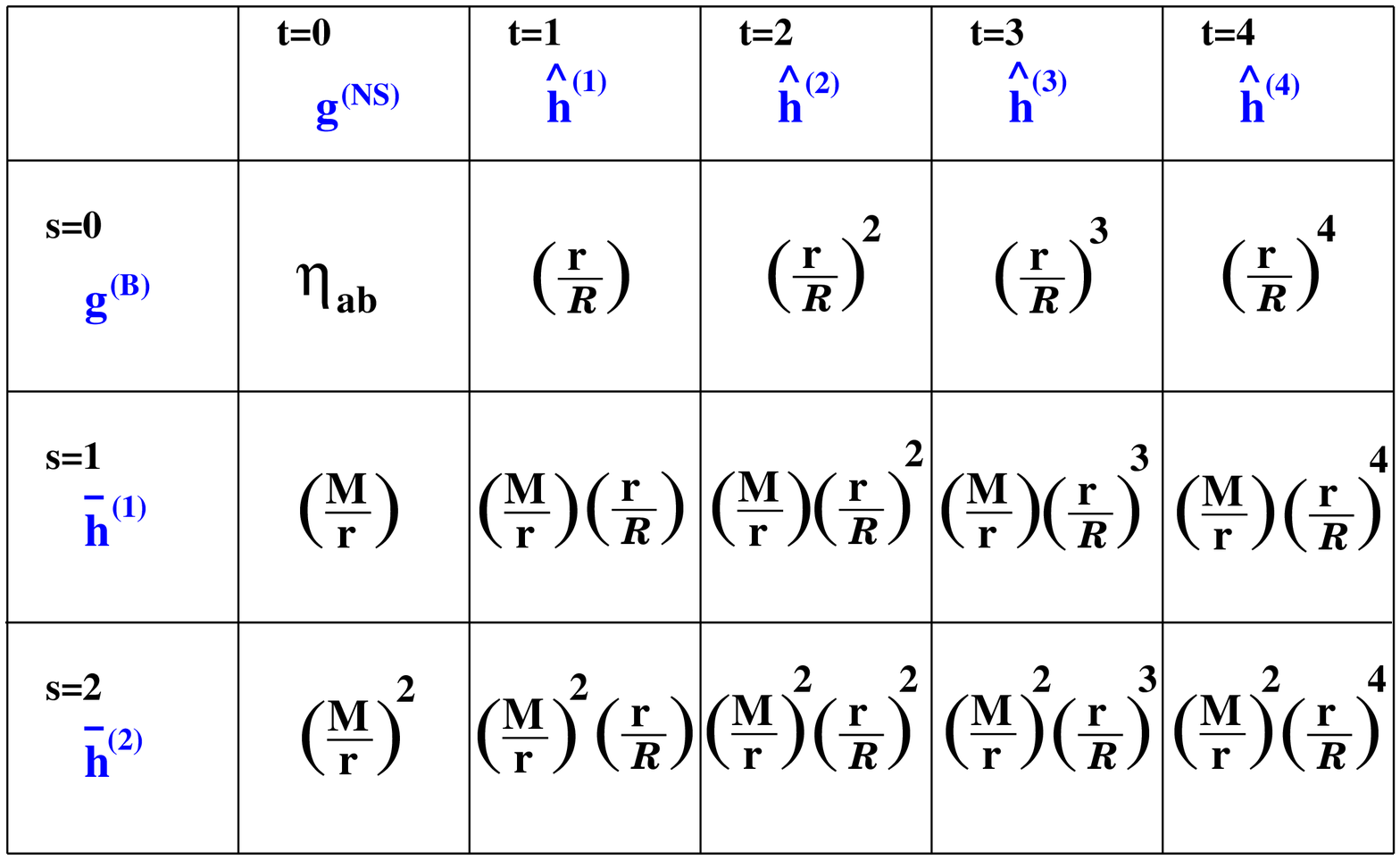,height=6cm,width=8.5cm,angle=0}}
\vskip 0.5cm
\caption{An illustration of the matching scheme for a single body
moving in a general vacuum exterior spacetime $g_{ab}^{(B)}$.
The tensors $g_{ab}^{({\rm NS})}$, ${\hat
h}_{ab}^{(1)}$, ${\hat h}_{ab}^{(2)}, \ldots$ when expanded in power
series in $M/r$ yield the columns in this diagram.  Similarly, the
the tensors $g_{ab}^{(B)}$, ${\bar h}_{ab}^{(1)}$, ${\bar h}_{ab}^{(2)}$
when expanded in power series in $r/{\cal R}$ and when acted on by the
pullback map \protect{(\ref{varphidef})} yield the rows.
The matching procedure consists of demanding consistency between the two
sets of expansions at each order $(s,t)$ in the double expansion,
wherein the metric perturbations scale as $(M/r)^s (r / {\cal R})^t$.
In the external spacetime, the
metric perturbation ${\bar h}_{ab}^{(1)}$ is determined by the $(s,t)
= (1,0)$ matching, and ${\bar h}_{ab}^{(2)}$ by the $(s,t) = (2,0)$
and $(2,1)$ matchings.  In the internal scheme, the metric
perturbation ${\hat h}_{ab}^{(1)}$ vanishes identically,
${\hat h}_{ab}^{(2)}$ is determined by the $(s,t) =
(0,2)$ matching, ${\hat h}_{ab}^{(3)}$ is determined by the $(0,3)$
and $(1,3)$ matchings, and ${\hat h}_{ab}^{(4)}$ by the $(0,4)$,
$(1,4)$ and $(2,4)$ matchings.
}
\label{fig:matching}
\end{figure}
{\vskip 0.25cm}

Turn next to the perturbations ${\hat h}_{ab}^{(3)}$ and ${\hat
h}_{ab}^{(4)}$.  The free parameters in these perturbations are the
quantities $b_{Jm}^{(3)}$, $d_{Jm}^{(3)}$, $b_{Jm}^{(4)}$ and
$d_{Jm}^{(4)}$ in Eqs.\ (\ref{hath4-general-1}) --
(\ref{T4-general-1}), and are determined by the matching scheme.  As
explained in Sec.\ \ref{response:centraldensity} above, the values of these
parameters will not affect the change in central density of the
neutron star and thus are unimportant for our purposes.  For
completeness we briefly mention how these parameters are determined.
First, since ${\hat h}_{ab}^{(3)}$ diverges like $r^3$ at large $r$,
it contains only $J=3$ and $J=2$ pieces.  The $J=3$ piece is
determined by the $(s,t) = (0,3)$ matching, and depends linearly on the
spatial derivatives of the electric and magnetic parts of the Weyl
tensor of the background spacetime $g_{ab}^{(B)}$, evaluated on the
geodesic.  The $J=2$ piece is determined by the $(1,3)$ matching, and
depends on the first order perturbation ${\bar h}_{ab}^{(1)}$ in the
external spacetime as well as on the background metric
$g_{ab}^{(B)}$.  Specifically, ${\bar h}_{ab}^{(1)}$ is first
determined by the $(1,0)$ matching \cite{Mino}, and from this one can
calculate the $(1,3)$ element of the matrix in Fig.\
\ref{fig:matching}, and hence infer the $J=2$ piece of ${\hat
h}_{ab}^{(3)}$.  Note that $h_{ab}^{(3)}$ thus depends on the geometry
of $g_{ab}^{(B)}$ not just in a neighborhood of the geodesic, but also
non-locally \cite{tailmetric}.  In a similar way ${\hat h}_{ab}^{(4)}$
contains pieces
with $0 \le J \le 4$ in the term ${}^0 h_{ab}^{(4)}$ in Eq.\
(\ref{hath4-general-1}) which are independent of the $t=4$ matchings,
and pieces with $2 \le J \le 4$ in the second term in
Eq.\ (\ref{hath4-general-1}) which are determined by the $t=4$
matchings.  These additional $2 \le J \le 4$ pieces depend on ${\bar
h}_{ab}^{(1)}$ and ${\bar h}_{ab}^{(2)}$ in addition to $g_{ab}^{(B)}$
and are determined by the $(0,4)$, $(1,4)$ and $(2,4)$ matchings.

Returning to the $O(\varepsilon^2)$ perturbation, it follows from the
arguments of Sec.\ \ref{response:centraldensity} that the leading
order change
(\ref{rhocbasic1}) in central density depends quadratically on ${\hat
h}_{ab}^{(2)}$, and hence quadratically on ${\cal E}_{ij}$ and ${\cal
B}_{ij}$, the curvatures of the external background spacetime
$g_{ab}^{(B)}$ evaluated on the worldline.  Furthermore, since all the
relevant equations are elliptic,
the dependence of $\delta \rho_c$ on ${\cal E}_{ij}$ and ${\cal
B}_{ij}$ is local in time.  Hence, invariance under rotations yields
\begin{equation}
{\delta \rho_c(t) \over \rho_c} = c_1 {\cal E}_{ij}(t) {\cal E}^{ij}(t) + c_2
{\cal B}_{ij}(t) {\cal B}^{ij}(t) + O({\cal R}^{-5}),
\label{finalans0}
\end{equation}
where $c_1$ and $c_2$ are constants of dimension (length)$^{-4}$ that
depend on $M$, $R$ and on the equation of state.  (A cross term
between the ${\cal E}_{ij}$ and ${\cal B}_{ij}$ fields is forbidden by
parity arguments.)

Thus, Eqs. (\ref{fluid2}) -- (\ref{fluid4}) have the same structure
as their Newtonian and post-Newtonian counterparts.  Equations
(\ref{fluid2}), (\ref{fluid3}) together with Eqs.~(\ref{hB2}),
(\ref{hB3}) describe distortions of the star induced by the external
tidal fields, where there is no change in the star's central density
and no excitation of the star's $J=m=0$
modes.  Equation (\ref{fluid4}) describes the second-order effect of
the leading-order
tidal field ${\hat h}_{ab}^{(2)}$ on the star's structure.  Exactly as
in Newtonian gravity, it is this second-order effect of the leading-order
tidal field that excites the spherically symmetric, radial modes of
the star and changes the star's central density and angle-averaged
radius.

It is possible to understand this result in a fairly simple,
intuitive way.  Consider first a weakly self-gravitating body in an
external gravitational field in the tidal limit $R \ll {\cal R}$.
Simply analyzing the dynamics of the test body using Fermi-normal
coordinates (\ref{background0}) allows one to immediately conclude
that the effect of the external field on the body's internal dynamics
must scale as ${\cal R}^{-2}$; this is just the equivalence principle.
The fact that there is no spherically symmetric interaction at this
order (that is, that the interaction can be called a ``tidal''
interaction) follows from algebraic properties of general relativity
-- the relativistic generalization of the familiar fact that the trace
of the Newtonian tidal force tensor $\partial^2 \Phi / \partial x^i
\partial x^j$ vanishes.  As a consequence of this vanishing of the
spherically symmetric interaction, all radial crushing or
anti-crushing forces must scale as ${\cal R}^{-4}$.
At first sight the above argument does not apply to a strongly
self-gravitating body.  However, the essence of the argument can be
carried
through.  What is relevant for determining the interaction between the
body and the external field are Einstein's equations in the matching
region $R \ll r \ll {\cal R}$  (the body's ``local asymptotic rest
frame'' \cite{kip}).  The asymptotic value of the external spacetime's
curvature tensors in this region (which is the region $r \to 0$ as
seen in the external spacetime) act as a source for the interaction,
and their scaling ($\propto {\cal R}^{-2}$) and algebraic properties
determine the nature of the interaction in the same way as for a
weakly self-gravitating body.

To conclude, we have shown that, just as in Newtonian gravity, {\it the
leading order change in the central density of a fully relativistic
spherical star freely falling in an external vacuum gravitational
field is given by the star's second-order response to the
leading order, external tidal field}.

\section{MODIFIED MATCHED ASYMPTOTIC EXPANSION METHOD APPLICABLE TO
NEUTRON STARS IN A BINARY}
\label{sec:PN}

The analysis of Secs.\ \ref{sec:main} -- \ref{sec:response} assumes
that the spacetime outside the body is vacuum.  This assumption
entered in the equation of motion (\ref{externaleqns1}) satisfied by
the metric perturbation ${\bar h}_{ab}^{(1)}$, which is used to derive
the fact that the star travels along a geodesic of the background
metric $g_{ab}^{(B)}$ to leading order [cf.\ Eq.\ (\ref{centralworldline})
above].  It is possible to
modify the analysis of Secs.\ \ref{sec:main} -- \ref{sec:response} to
accommodate two freely falling bodies, for examples two neutron stars
moving in the vicinity of a supermassive black hole.  In this case the
metric perturbation ${\bar h}_{ab}^{(1)}$ describes the linearized
gravitational interactions of the two neutron stars, and one must
solve simultaneously for the external metric perturbations, for two
sets of internal metric perturbations, one for each star, and likewise
for two sets of embedding functions.  The scheme allows one to derive,
for example, the equations of motion of two ``point particles''
interacting via their linearized gravitational fields.

Such a calculational scheme is applicable in principle to an isolated
neutron star binary, but is poorly adapted to that situation. Since
the background metric $g_{ab}^{(B)}$ is a flat Minkowski
metric in this context, the calculations of Secs.\ \ref{sec:main} --
\ref{sec:response} of the leading order change in central density are
not applicable.  Moreover, to achieve our goal of determining
the scaling of the change in central density with the parameters
$\epsilon$ and $\alpha$ discussed in Sec.\ \ref{intro:coupling}, one
should describe
the gravitational interactions of the two neutron stars not by metric
perturbations ${\bar h}_{ab}^{(1)}$ and ${\bar h}_{ab}^{(2)}$, but
rather in terms of a post-Newtonian expansion.  For these reasons, in
this section we outline a modified, matched asymptotic expansion
calculational method in which metric perturbations in an internal
scheme are matched onto post-Newtonian quantities in an external scheme.
The modified method will allow us to calculate the change in central
density for neutron stars in a binary.

As in Sec.\ \ref{sec:main} above, the method consists of an internal
scheme, an external scheme, and a matching scheme.  A key difference
is that there are two sets of internal schemes and two sets of
matchings, one for each star, all of which must be solved
self-consistently.

\subsection{The external scheme}
\label{PN:externalscheme}

Let $M^{(B)}$ be the external manifold in which the neutron stars
move.  The gravitational field is described in the external scheme by
the standard post-Newtonian expansion in vacuum.  Thus the zeroth order,
background solution is just a Newtonian spacetime, instead of the
vacuum Lorentzian metric $g_{ab}^{(B)}$ we had previously.
The metric perturbations ${\bar h}_{ab}^{(s)}$ are replaced by
post-Newtonian fields, post-post-Newtonian fields, etc
\cite{notePNGW}.

Now, any Newtonian spacetime can be characterized by a lengthscale ${\cal
L}$ and a massscale $M$, such that the typical value of the Newtonian
potential is $\sim M/{\cal L}$ and such that the local radius of
curvature ${\cal R}_c$ is given by ${\cal R}_c^{-2} \sim M /
{\cal L}^3$.  In our example of a neutron star binary, ${\cal L}$ will
be just the orbital separation $L$.  The post-$s/2$-Newtonian
fields, for $s = 0,1,2 \ldots$, scale as $M^{s/2}$, but have
no definite scaling with respect to ${\cal L}$.  This is analogous to
the behavior of the expansion (\ref{exteriormetric1}).  As is well
known, the post-Newtonian fields corresponding to odd values of $s$
vanish identically for $s = 1,3$ and start at $s=5$.

In a suitable coordinate system ${\bar x}^\alpha = ({\bar t},{\bar x}^i)$,
the metric up to
post-1-Newtonian order can be written in the standard form \cite{Will}
\begin{eqnarray}
ds^2 &=& - \big[1 + 2 \varepsilon^2 \Phi({\bar x}^i,\varepsilon {\bar t})
+ 2 \varepsilon^4 \Phi({\bar x}^i,\varepsilon {\bar t})^2 + 2
\varepsilon^4 \Psi({\bar x}^i,\varepsilon {\bar t})
\nonumber \\
\mbox{} && +O(\varepsilon^6)\big] d{\bar t}^2  + 2 d{\bar x}^i d{\bar
t} \left[ \varepsilon^3 W_i({\bar x}^i,\varepsilon {\bar t}) +
O(\varepsilon^5) \right]
\nonumber \\
\mbox{} && + d{\bar x}^i d{\bar x}^j \left[ \delta_{ij} - 2
\varepsilon^2 \Phi({\bar x}^i,\varepsilon {\bar t}) \delta_{ij} +
O(\varepsilon^4) \right],
\end{eqnarray}
where $\varepsilon \ \propto \ \sqrt{M}$ is a formal expansion
parameter.  Equivalently, by
making a gauge change ${\bar t} \to {\bar t} / \varepsilon$, the
metric can be written as
\begin{eqnarray}
ds^2 &=& - {1 \over \varepsilon^2} \Big[1 + 2 \varepsilon^2 \Phi({\bar
x}^i,{\bar t}) + 2 \varepsilon^4
\Phi({\bar x}^i,{\bar t})^2 + 2 \varepsilon^4 \Psi({\bar x}^i,{\bar t})
\nonumber \\
\mbox{} && +O(\varepsilon^6)\Big] d{\bar t}^2  + 2 d{\bar x}^i d{\bar
t} \left[ \varepsilon^2 W_i({\bar x}^i,{\bar t}) + O(\varepsilon^4) \right]
\nonumber \\
\mbox{} && + d{\bar x}^i d{\bar x}^j \left[ \delta_{ij} - 2
\varepsilon^2 \Phi({\bar x}^i,{\bar t}) \delta_{ij} + O(\varepsilon^4) \right].
\label{pn-metric1}
\end{eqnarray}
In vacuum one can pick a gauge in which the potentials $\Phi$, $\Psi$ and
$W_i$ obey the equations \cite{Will}
\beq
\nabla^2 \Phi = \nabla^2 W_i = \nabla^2 \Psi + {\ddot \Phi} = 0,
\endeq
where $\nabla^2$ is the Laplacian of flat space.

The post-Newtonian expansion can also be described in a coordinate-free
way \cite{Dautcourt}.  Let $g^{ab}(\varepsilon)$ be a one parameter
family of vacuum metrics which are $C^2$ in $\varepsilon^2$ in a
neighborhood of $\varepsilon=0$, such that the limit $\varepsilon \to
0$ of $g^{ab}(\varepsilon)$ exists and is of signature $(0,+,+,+)$.
[The formula (\ref{pn-metric1}) gives an approximate version of
such a family].
Then there exist tensor fields $h^{ab}$, $t_a$ and a connection $D_a$
such that
\begin{eqnarray}
\label{pn-1}
g^{ab}(\varepsilon) &=& h^{ab} + O(\varepsilon^2)  \\
\varepsilon^2 g_{ab}(\varepsilon) &=& - t_a t_b + O(\varepsilon^2)\\  
\nabla_a(\varepsilon) &=& D_a + O(\varepsilon^2).
\label{pn-3}
\end{eqnarray}
The quantities $D_a$, $h^{ab}$ and $t_a$ comprise a Newtonian
spacetime.  In Newtonian or ``inertial'' coordinate systems $({\bar
x}^i,{\bar t})$, these quantities are given
by $t_a = (d{\bar t})_a$, $h^{ab} = \delta_{ij} (\partial / \partial
{\bar x}^i)^a (\partial / \partial {\bar x}^j)^b$, and $D_a$ is given
in terms of the Newtonian potential $\Phi$
by the only non-vanishing connection
coefficient being $\Gamma^i_{tt} =  \Phi_{,i}$.
The Newtonian fields satisfy the additional relations
\beq
D_a t_b = D_a h^{bc} = h^{ab} t_b = h^{a[b} R^{c]}_{\ (de)a}=0,
\label{Newtonian-identities}
\endeq
where $R^a_{\ bcd}$ is the curvature of the connection $D_a$.
The last of the relations (\ref{Newtonian-identities}) is just the
the limit $\varepsilon \to 0$ of the identity
$g^{a[b}(\varepsilon) R^{c]}_{\ (de)a}(\varepsilon)=0$.  The higher
order correction terms in Eqs.\ (\ref{pn-1}) -- (\ref{pn-3}) of
order $O(\varepsilon^s)$ collectively describe
the post-$s/2$-Newtonian fields; we shall not need the precise forms
of these fields here \cite{Tichy}.

\subsection{The internal scheme}
\label{PN:internalscheme}

The internal scheme is very similar to that described in Sec.\
\ref{internalscheme} above.
The interior metric still is given by
Eq.\ (\ref{interiormetric}), however now each term $\varepsilon^t {\hat
h}_{ab}^{(t)}$ in that expansion scales as ${\cal L}^{-t/2}$ rather than
${\cal R}^{-t}$, and has no definite scaling with respect to $M$.
The reason that we need to include half-integral powers of $1/{\cal L}$ is
that time derivatives will scale like ${\cal L}^{-3/2}$.

The equations of motion (\ref{fluid2}) -- (\ref{hB4}) are modified by
the following two considerations.
First, time-derivatives now scale as $\varepsilon^3$ rather than
$\varepsilon$, which modifies the form of the perturbation analysis.
Hence, a time derivative of the field ${\hat h}_{ab}^{(t)}$ will enter
as a source term in the equation for the field ${\hat
h}_{ab}^{(t+3)}$.  The perturbations ${\hat h}_{ab}^{(t)}$ with $t$
odd are due entirely to such time-derivative source terms.
Since it turns out that the first non-vanishing perturbation occurs at
$O(\varepsilon^6)$, the first non-vanishing ${\hat h}_{ab}^{(t)}$ with
$t$ odd occurs for $t=9$.  The second consideration is that we need
now to consider the
perturbations up to order $t=12$ instead of up to order $t=4$.
This is not as complex as it seems since it suffices to consider even
values of $t$ until $t=9$.  In addition, as already mentioned the
first non-vanishing perturbation is ${\hat h}_{ab}^{(6)}$.

The structure of the resulting equations of motion is closely
analogous to that of Eqs.\ (\ref{fluid2}) -- (\ref{hB4}), and the
general solutions follow the same pattern as the solutions
(\ref{linear-solns-h}), (\ref{linear-solns-T}) and
(\ref{hath3-general-1}) -- (\ref{T4-general-1}).  As before, the
crucial aspect
of the solutions that we shall use is that
the first non-vanishing perturbation in the $J=0$ sector cannot arise
directly from matching to the external scheme, but must arise from a
source term that is quadratic in a lower order field.

In what follows, we neglect entirely odd values of $t$.  This is
justified since, just as in Sec.\ \ref{response:matching} above, all
time-derivative-generated terms have no $J=0$ parts and will not
contribute to the leading order change in central density.

\subsection{The matching scheme}
\label{PN:matchingscheme}

The construction of the matching scheme parallels that given in Sec.\
\ref{matchingscheme} above.  One needs to specify an embedding
\begin{equation}
\varphi: \ M^{({\rm NS})} \to M^{({\rm B})},
\label{em1}
\end{equation}
of the neutron star spacetime into the external spacetime, i.e.,
a mapping between the asymptotically Lorentzian
coordinates $x^\alpha$ of $M^{({\rm NS})}$ and the coordinates
${\bar x}^\alpha$ of $M^{({\rm B})}$.
We can write this mapping as [compare Eq.\ (\ref{varphidef}) above]
\begin{eqnarray}
{\bar x}^\alpha &=& {\bar x}^\alpha(t,x^i) \nonumber \\
\mbox{} &=&
\sum_{p=0} \, \left( { M \over {\cal L} } \right)^{p/2} \, \sum_{r=0} \,
\sum_{I_r} \ \ {}^{(r,p)}{\cal F}^\alpha_{I_r}(t) \, x^{I_r},
\label{varphidef1}
\end{eqnarray}
where $I_r$ denotes the multi-index $(i_1, \ldots, i_r)$ and $x^{I_r} \equiv
x^{i_1} x^{i_2} \ldots x^{i_r}$.  Equation (\ref{varphidef}) is a
Taylor expansion of $\varphi$ in terms both of the spatial coordinates
$x^i$ at each fixed $t$, and also in terms of the parameter $\sqrt{M / {\cal
L}}$.  The terms with $p$ odd are vanishing for $p=1,3$; the first
non-vanishing terms with $p$ odd start at $p=5$.
Thus, the worldline of the center of
the neutron star gets mapped onto the worldline
\begin{equation}
{\bar x}^\alpha(t) = {}^{(0,0)}{\cal F}^\alpha(t) + {M \over {\cal L}}
{}^{(0,2)}{\cal F}^\alpha(t) + O[(M/{\cal L})^2].
\label{centralworldline1}
\end{equation}
The matching procedure described below can be used to show that the
first term in Eq.\ (\ref{centralworldline1}) represents a worldline
satisfying Newtonian equations of motion, that the second term is the
first post-Newtonian, point mass correction, etc.

Next, from the interior metric $g_{ab}^{({\rm INTERIOR})}$ on
$M^{({\rm NS})}$ one construct the metric
\beq
{\bar g}_{ab} = \left( \varphi^{-1} \right)_* \chi_* g_{ab}^{({\rm
INTERIOR})}
\label{pullback1}
\endeq
on $M^{(B)}$.  Here $\chi : M^{({\rm NS})} \to M^{({\rm NS})}$ is the
mapping defined after Eq.\ (\ref{Lambdadef}) above, with the parameter
$\varepsilon$ chosen to have the value $\sqrt{M}$.  The metric $\chi_*
g_{ab}^{({\rm INTERIOR})}$ will have a dependence on $\sqrt{M}$ like
that of the metric (\ref{pn-metric1}) has on $\varepsilon$.
Now perform a double power series expansion of the metric ${\bar
g}_{ab}$, its inverse ${\bar g}^{ab}$ and its associated connection
${\bar \nabla}_a$ in terms of the parameters $M/r$ and $r/{\cal L}$,
as in Eq.\ (\ref{doubleexpansion}) above.  These double power series
expansion must be consistent, order by order, with expansions in
powers of $r / {\cal L}$ of the post-$s/2$-Newtonian fields for $s = 0,
2, 4, 5, \ldots$ of the external scheme.
As before, demanding such consistency determines both the embedding free
functions and the appropriate solutions in the internal and external
schemes, up to some gauge freedom.

{\vskip 0.5cm}
\begin{figure}
{\psfig{file=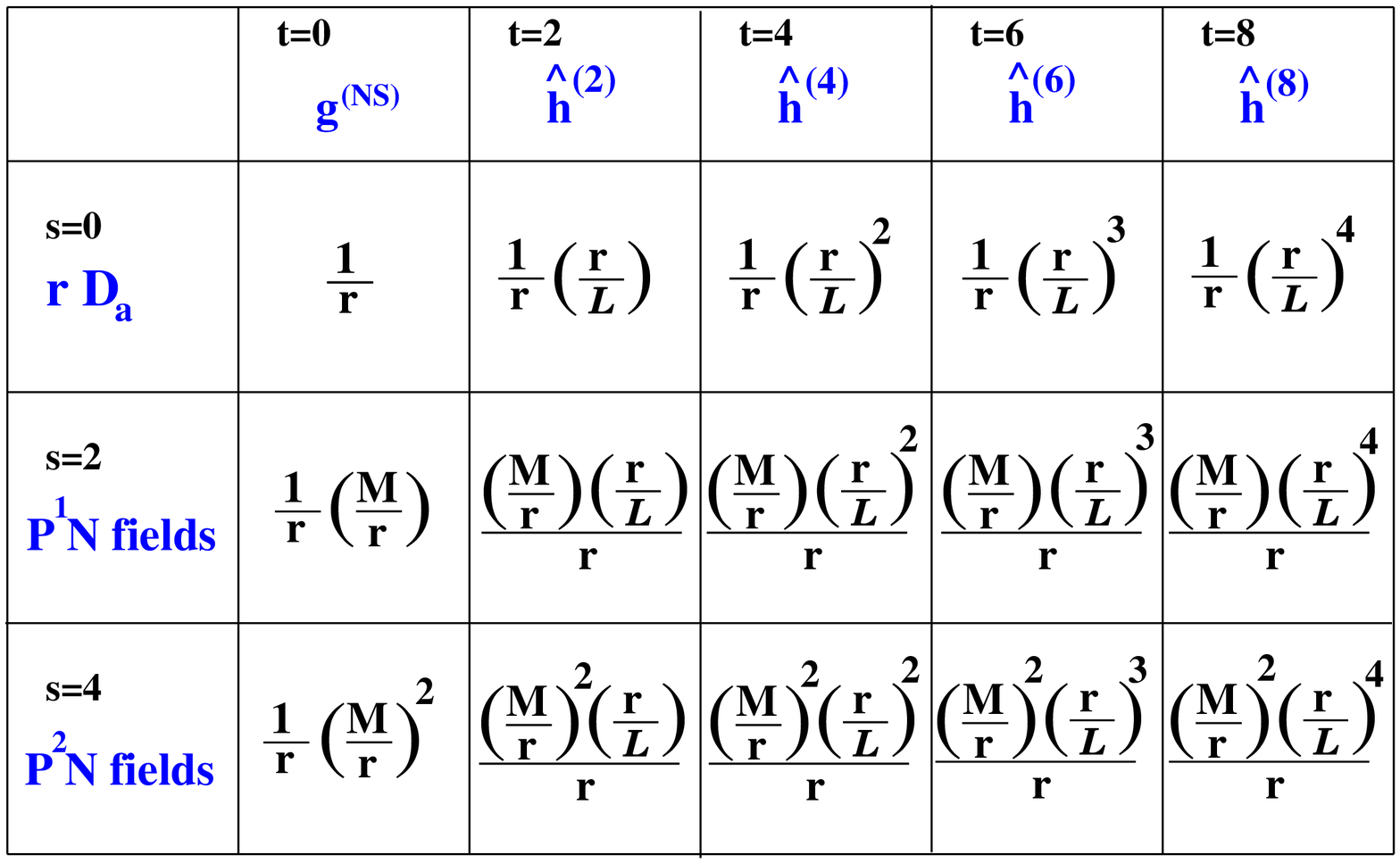,height=6cm,width=8.5cm,angle=0}}
\vskip 0.5cm
\caption{An illustration of the matching scheme for a body
moving in a spacetime described by a post-Newtonian approximation.
In the external spacetime, the post-$s/2$-Newtonian fields for $s=0,2,4$,
when expanded in power series
in $r/{\cal L}$, yield the rows in this diagram.
The sum of $g_{ab}^{({\rm NS})}$, ${\hat h}_{ab}^{(1)}$, ${\hat
h}_{ab}^{(2)}, \ldots$ when acted on by the
pullback map (\ref{pullback1}) yields a tensor field ${\bar
g}_{ab}(M)$ on the external spacetime.  When one considers the triple
$\left(M \, {\bar g}_{ab}(M), {\bar g}^{ab}(M), {\bar \nabla}_a(M)\right)$, and
expands this triple order by order in $M$, one obtains at $O(M^{s/2})$
the post-$s/2$-Newtonian fields, for $s = 0,2,4,5 \ldots$.
The expansions in powers of $r / {\cal L}$ of these fields must agree
with those of the external spacetime post-Newtonian fields.
In the internal scheme, the metric
perturbations ${\hat h}_{ab}^{(2)}$
and ${\hat h}_{ab}^{(4)}$ vanish identically, and ${\hat
h}_{ab}^{(6)}$ is determined by the $(s,t) = (0,6)$ matching.  The
first change in central density occurs in ${\hat h}_{ab}^{(12)}$.
}
\label{fig:matching1}
\end{figure}
{\vskip 0.25cm}

Consider for example the case $s=0$, which is just the matching onto a
Newtonian spacetime.  The
limit $M \to 0$ of the fields ${\bar g}_{ab}$, ${\bar g}^{ab}$ and ${\bar
\nabla}_a$ yields yields Newtonian fields ${\bar h}^{ab}$, ${\bar
t}_a$ and ${\bar D}_a$, cf. Eqs.\ (\ref{pn-1}) -- (\ref{pn-3}) above.
The expansion in powers of $r / {\cal L}$ of these fields must agree,
order by order, with the expansions in powers of $r / {\cal L}$ of the
Newtonian fields $h^{ab}$, $t_a$, and $D_a$ of the external scheme.
If we choose the coordinate system (\ref{neutronstar1}) in the
internal scheme and a Newtonian or inertial coordinate system $({\bar
x}^i, {\bar t})$ in the internal scheme, then the fields $h^{ab}$,
${\bar h}^{ab}$ and $t_a$, ${\bar t}_a$ will coincide to all orders in
$r/{\cal L}$ if we choose the embedding (\ref{em1}) to be of the form
${\bar t}(x^j,t) = t$, ${\bar x}^i(x^j,t) = x^i + f^i(t)$.  Then one
just has to match the Newtonian potentials order by order in $r /
{\cal L}$.  

This matching to zeroth order in $r/{\cal L}$ [i.e., the $(s,t) = (0,0)$ 
matching] yields the usual Newtonian point-mass equations of motion.
Hence, when one solves for the embedding functions of both stars and
for the interior and exterior gravitational fields up to order $(s,t) =
(0,0)$, the result is two static spherical neutron stars moving along
their Newtonian orbits.  This then serves as the starting point for
calculating higher order perturbations.

This matching procedure is illustrated schematically in Fig.\
\ref{fig:matching1}.  Our arguments below will be independent of the
details of the post-Newtonian and post-post-Newtonian matchings, so we
do not need to describe those here.

\subsection{Change in central density}

The leading order change in central density $\delta \rho_c$ of a
neutron star in a binary can now be deduced by arguments analogous to
those given in Sec.\ \ref{response:matching} above.  First, in the internal
scheme we need consider only even values of $t$, as
explained in Sec.\ \ref{PN:internalscheme}.  Next, consider the
perturbations ${\hat h}_{ab}^{(2)}$ and ${\hat h}_{ab}^{(4)}$.  If
these perturbations were to be non-zero (or not pure gauge), then they
would need to diverge at large $r$ like $r^J$ for some $J \ge 2$, from
the analysis in Sec.\ \ref{sec:internal-scheme}.  But
from Fig.\ \ref{fig:matching1}, ${\hat h}_{ab}^{(2)} \sim {\rm const}$
and ${\hat h}_{ab}^{(4)} \ \propto \ r$ at large $r$. Hence the first
non-vanishing perturbation is ${\hat h}_{ab}^{(6)}$, which is purely
$J=2$.  Note that this perturbation is determined entirely by
the exterior Newtonian fields from the $(s,t) = (0,6)$
matching.

Since the first non-vanishing perturbation is ${\hat h}_{ab}^{(6)}$, 
the first change in central density must scale as $({\hat
h}_{ab}^{(6)})^2$ by the same arguments as before, so we must have
\beq
{\delta \rho_c \over \rho_c} \propto { 1 \over {\cal L}^6}
\label{scale0}
\endeq
at large ${\cal L}$.  The constant of proportionality in
Eq.~(\ref{scale0}) must have dimensions of $({\rm
length})^6$, and the only relevant dimensionful parameters are the
mass $M$ of the star and its radius $R$.  We can
therefore write
\begin{equation}
{\delta \rho_c  \over \rho_c} = F[M/R] \left({R\over {\cal
L}}\right)^6 = F(\epsilon)
\alpha^6 + O(\alpha^7),
\label{sc1}
\end{equation}
where $F$ is a dimensionless function of a dimensionless variable, and
we have used the dimensionless variables  $\alpha \equiv R/L$ and
$\epsilon \equiv M/R$ discussed in the Introduction.  The function $F$
will depend on the equation of state.

\section{DISCUSSION AND CONCLUSIONS}
\label{sec:conclusions}

As discussed in the Introduction, the result (\ref{sc1}) is in
disagreement with Refs.~\cite{wmm,ReturnOfTheJedi} in the regime
$\alpha \ll 1$ where there should be agreement, since the scaling
found in the numerical simulations is $\delta \rho_c / \rho_c \propto
\alpha^2$ as $\alpha \to 0$.  This disagreement in scaling at large
orbital separations is strong evidence that the star crushing
effect seen in Refs.\ \cite{wmm,ReturnOfTheJedi} is not physical.

In addition, at the location of the instability seen by Wilson et.\
al.\ , the dimensionless parameter $\alpha^6$ is of order $10^{-4}$,
so one might expect the perturbative result (\ref{sc1}) to be a good
approximation.  For $F(\epsilon)$ of order unity, the fractional
change in central density (\ref{sc1}) at this location is $\sim 1000$
times smaller than that seen in the numerical simulations
\cite{wmm,ReturnOfTheJedi}.

The analysis of this paper does not determine the sign of the function
$F(\varepsilon)$.  However, we can make the following argument.  Let
us expand this function as a power series, which yields an expression
of the form
\begin{equation}
{\delta \rho_c  \over \rho_c} = F_0 \alpha^6 \left[ 1 + F_1 \epsilon +
F_2 \epsilon^2 + \ldots \right],
\label{finalans}
\end{equation}
for some constant coefficients $F_0$, $F_1$, $F_2$ etc.
Now, it is known from Lai's Newtonian analysis that the
coefficient $F_0$ is negative \cite{lai,commentlai}.  Therefore, unless the
dimensionless coefficients $F_1$, $F_2$ etc are negative and large
($\sim -7$ for $F_1$), which seems unlikely, the change in central
density due to the tidal interaction will be negative.  Thus, the
stars angle-averaged radius will increase and the star will be more
stable and not less stable to radial collapse.

Note that our analysis assumes that the neutron stars are not spinning
in their local rest frames, although the dragging of inertial frames
means that
they will spin slightly with respect to distant stars; this effect is
incorporated in the choice of Fermi-normal coordinates in
Eq. (\ref{background0}).

Note also that our analysis assumes that the external tidal
fields are slowly varying compared to the dynamical timescales of the
neutron star, so that the stellar modes adiabatically follow their
driving forces.  This follows from the assumption $\alpha \ll 1$
underlying the perturbation expansion.  Clearly this adiabatic approximation
breaks down at $\alpha \sim 1$.  However, one can ask
how well we might expect the adiabatic approximation to be working
at the location of the instability.
For the neutron star $f$-modes and $p$-modes, the idealization is a good
approximation:  From the point of view of the
companion star, the timescale over which the external tidal fields is
varying is ${\cal T} \sim \sqrt{L^3/M}$.  This
timescale is long compared to the internal dynamical time of the
neutron star (the characteristic timescale of the $f$-modes) $\sim
\sqrt{R^3/M}$.  Hence, these modes will equilibrate rapidly in
response to the tidal perturbations and the approximation is
fairly good.   The $g$-modes of the neutron star on the other hand have
frequencies $\ll \sqrt{M/R^3}$ and are resonantly excited \cite{Dong1,Kokkotas}
in the Newtonian approximation near the end of the inspiral; for these
modes the approximation is inappropriate.  This is a limitation of the
applicability of our analysis.  However, it seems unlikely
that resonant $g$-mode excitations could be responsible for the star crushing
effect seen in the Wilson-Mathews-Marronetti simulations, since the
predicted amplitudes of excitation in the Newtonian approximation are
fairly small \cite{Dong1,Kokkotas}.

Finally, consider the implications of our result for the gravitational
wave
signature of the inspiral.  Previous analyses
\cite{BildstenCutler,Kochanek,lai,LaiWiseman}
have shown that to Newtonian order, the effect of the finite size of
the neutron stars on the gravitational waveform is small, and in
fact is negligible for
the purposes of signal detection.  For the initial LIGO
interferometers, $\sim 95 \%$ of the signal-to-noise ratio will have
been accumulated before the gravitational wave frequency reaches $400
\, {\rm Hz}$ [see, e.g., Eq. (2.23) of Ref.~\cite{CF}].  The
total accumulated phase error in the signal predicted by Newtonian
tidal interactions by $400 \, {\rm Hz}$ is $\ll 1$; see Eq.~(7.5) of
Ref.~\cite{lai} and Ref.~\cite{LaiWiseman}.  We have shown using a
fully relativistic treatment of tidal interactions that,
as one would expect, the Newtonian predictions for tidal interactions
are valid except for correction terms of order $M/R$ and $M/L$, that
is, fractional corrections $\alt 1$.
Therefore, the effects of tidal interactions and of the finite size of
the neutron stars will be unimportant for signal detection.

As this paper was being completed, we learned of a similar analysis by
Thorne \cite{kip}.  Thorne shows that the change in central density is
$\propto \alpha^6$, but also that it is always negative, stabilizing
the neutron star.

\acknowledgments
The author thanks Scott Hughes, James Lombardi, Grant Matthews, Ted
Quinn, Saul Teukolsky, Kip Thorne, Ira Wasserman and Alan Wiseman for
useful discussions, and David Chernoff for useful discussions and for
detailed comments on the manuscript.  This research was supported in
part by NSF grant PHY--9722189 and by a Sloan Foundation fellowship.

\newpage
\onecolumn

\begin{table}
\caption{A summary of the orders at which various physical effects
occur.  The parameter $\alpha$ is the tidal expansion parameter
(stellar radius)/(orbital separation), and $\epsilon$ is the strength
of internal gravity in each star, $\epsilon = $(stellar mass)/(stellar
radius).  A perturbation to the orbital motion is said to scale as
$\alpha^n \epsilon^m$ if the ratio of the perturbing force to the
Newtonian orbital force is of this order.  Similarly, the distortion
of each star is said to be of order $\alpha^n \epsilon^m$ if the
dimensionless measure of distortion $(I_{xx} - I_{yy})/I_{xx}$ is of
this order.
Note that to post-1-Newtonian order, all the physical
effects scale the same way with $\alpha$ as their Newtonian
counterparts.  The pattern for the stellar modes is repeated at all
higher orders in $\epsilon$.  The entries in the orbital motion
columns can be deduced from the entries in the stellar modes columns;
if $\delta {\cal E}$ is the energy transferred between the
stellar modes and the orbital motion, then the fractional change in
the orbital motion is $\sim L \delta {\cal E} / M^2$, while the
fractional distortion of the star is $\sim R \delta {\cal E} / M^2$,
one power of $\alpha$ smaller.
\label{table1}}
\begin{tabular}{llllll}
&\mbox{\ \ \ \ \ \ \ \ \ \ \ \ \ \ \ \ \ Newtonian : $O(\epsilon^0)$}
&&&\mbox{\ \ \
\ \
\ \ \ \ \ \ \ \ \ \ \ \ \ \ \ \ \ \ \ \ \ \ \ \ \ $O(\epsilon)$} &\mbox{}\\
\tableline
\mbox{Tidal order}&\mbox{Orbital\ motion}&\mbox{Stellar\
modes}&\mbox{\ \ \ \ \ \ \ \ \ }&\mbox{Orbital\  motion}&\mbox{Stellar\ modes}
\\
\tableline
$O(\alpha^0)$ & \mbox{\ \ \ \ \ \ $\uparrow$} & \mbox{\ \ \ \ \ \ $\uparrow$}
&&
\mbox{Newtonian\ orbit} & \mbox{\ \ \ \ \ \ $\uparrow$} \\
$O(\alpha^1)$ & \mbox{\ \ \ \ \ \ $\vert$} & \mbox{static\ spherical\ star}
&& \mbox{\ \ \ \ \ \ $\uparrow$} & \mbox{static\ spherical\ star} \\
$O(\alpha^2)$ & \mbox{No\ change\ in\ orbit} & \mbox{\ \ \ \ \ \ $\downarrow$}
&& \mbox{post-1-Newtonian} & \mbox{\ \ \ \ \ \ $\downarrow$} \\
$O(\alpha^3)$ & \mbox{\ \ \ \ \ \ $\vert$} & \mbox{\ \ \ \ \ \ $\uparrow$} &&
\mbox{point\ particle\ orbit} & \mbox{\ \ \ \ \ \ $\uparrow$} \\
$O(\alpha^4)$ & \mbox{\ \ \ \ \ \ $\downarrow$} & \mbox{tidal deformation}
&& \mbox{\ \ \ \ \ \ $\downarrow$} & \mbox{post-1\ deformation}\\
$O(\alpha^5)$ & \mbox{\ \ \ \ \ \ $\uparrow$} & \mbox{\ \ \ \ \ \
$\downarrow$} && \mbox{\ \ \ \ \ \ $\uparrow$} & \mbox{\ \ \ \ \ \
$\downarrow$}\\
$O(\alpha^6)$ & \mbox{Orbit\ responds} & \mbox{\ \ \ \ \ \ $\uparrow$} &&
\mbox{orbit\ modified} & \mbox{\ \ \ \ \ \ $\uparrow$}\\
$O(\alpha^7)$ & \mbox{to tidal coupling} & \mbox{first\ change} &&
\mbox{\ \ \ \ \ \ $\vert$} & \mbox{post-1\ correction\ to}\\
$O(\alpha^7)$ & \mbox{\ \ \ \ \ \ $\vert$} & \mbox{in\ central\ density} &&
\mbox{\ \ \ \ \ \ $\vert$} & \mbox{change\ in\ central\ density}\\
$O(\alpha^8)$ & \mbox{\ \ \ \ \ \ $\vert$} & \mbox{\ \ \ \ \ \
$\vert$} && \mbox{\ \ \ \ \ \
$\vert$} & \mbox{\ \ \ \ \ \ $\vert$} \\
\end{tabular}
\end{table}


\begin{references}
\bibitem{note88} The normal modes of the neutron star will be damped
by gravitational wave emission and are usually called quasinormal
modes in the context of a fully relativistic treatment; for simplicity
we will call them normal modes in this paper. The gravitational wave
damping is never important for our purposes.
\bibitem{note10} The required $\Delta R/ R$ will be of
order the fractional difference in the neutron star mass (when
isolated) from the maximum, critical mass.
\bibitem{note3a}  The orbital instability is also predicted to occur at
a gravitational wave frequency of $\sim 300 \, {\rm Hz}$, close to the
frequency $\sim 200 \, {\rm Hz}$ where initial LIGO and VIRGO
interferometers are expected to accumulate most of their
detection signal-to-noise ratio.
\bibitem{note11} An exception is Ref.~\cite{Lombardi}, which analyzes
equilibria of binary neutron stars including the effect of the dominant
post-Newtonian terms but omitting post-Newtonian tidal terms.
\bibitem{extra} The arguments of this paper apply to an
inviscid isentropic fluid described by a one parameter equation of
state, $p = p(\rho)$, for which the 4 equations $\nabla_a T^{ab} =0$
determine the fluid's
evolution.  However, the arguments can easily be extended to the more
realistic case of a fluid described by a two parameter equation of
state $p = p(\rho,n)$, where $n$ is baryon number density, in which
case the additional equation $\nabla_a (n u^a)=0$ needs to be used.
\bibitem{note43} It is assumed that the time coordinate $t$ which is
used is such that the timelike killing field of of the static, neutron star
background metric is $\partial / \partial t$.
\bibitem{note6} In Eq.~(\ref{doubleexpansion}), the dependence on the
radius $R$ of the neutron star is not explicitly shown.  If it is
wished, the dependence can be made explicit by writing
$$
h_{ab}^{l,m} = \sum_{p=0}^l h_{ab}^{l,m,p} \left( {R \over M}
\right)^p,
$$
thus turning the expansion into a triple expansion in the parameters
$r/{\cal R}$, $M/r$ and $R/r$.  In a similar way, the expansion
(\ref{doubleexpansion}) can be understood to represent a simultaneous
expansion in the small quantities $r/{\cal R}_c$, $r / {\cal T}$, and
$r / {\cal L}$.
\bibitem{notelog}
In fact, explicit calculations show that the interior metric will have
a term which scales like
$$
\left( {M \over r} \right)^2 \ln \left({M \over r} \right) \,\left( {r
\over {\cal R}} \right)^4,
$$
generated by the third term on the right hand side of Eq.\
(\ref{hB4}) when $b_{2m}^{(2)} \ne 0$.  This term is a piece of ${\hat
h}_{ab}^{(4)}$, and matches onto a piece of the perturbation ${\bar
h}_{ab}^{(2)}$.  Thus for $t \ge 4$ it is
necessary to generalize the form
of the expansion (\ref{doubleexpansion}) to
$$
g_{ab}^{({\rm INTERIOR})} = \eta_{ab}
$$
$$
+ \sum_{s=0} \,
\sum_{t=0} \, \sum_{u=0} \, {\hat h}_{ab}^{(s,t,u)} \, \left( {M \over
r} \right)^s \, \left[ {M \over r} \ln \left({ M \over r} \right)
\right]^u \left( {r \over {\cal R}} \right)^t,
$$
and similarly for Eqs.\ (\ref{doubleexpansion1}) and
(\ref{doubleexpansion2}).  These modifications do not change the basic
calculational method.
\bibitem{gaugemethod}
This method of parameterizing gauge transformations beyond linear
order was developed in Ref.\ \cite{anec}.  Formulae for the gauge
transformations to all orders in perturbation theory are given in
Ref.\ \cite{extension}.
\bibitem{notelineart}
The expansions (\ref{texpand}) and (\ref{taexpand}) contain no terms
linear in $\varepsilon$.  This is not a restriction since such terms
could be set to zero by allowing a nonzero $\xi^{(1)\,a}$.
\bibitem{restriction}
This subsections analysis is valid for a Newtonian star moving in a
relativistic, external, vacuum gravitational field.  For a Newtonian
star moving in a Newtonian gravitational field, the analysis is
slightly modified; see Sec.\ \ref{PN:internalscheme}.
\bibitem{noteEB}
Note that with these definitions ${\cal B}_{ij}^{(2)}$ depends on the
metric perturbation ${\hat h}_{ab}^{(2)}$ as well as its associated
Riemann tensor perturbation, while ${\cal E}_{ij}^{(2)}$ depends only
on the Riemann tensor perturbation.  Also ${\cal B}_{ij}^{(2)}$ is
traceless with respect to $g_{ab}^{({\rm NS})}$ while ${\cal
E}_{ij}^{(2)}$ is not.
\bibitem{noteRMP} We use the symbols $L$ and $J$ instead of the
$l^\prime$ and $l$ of Ref.\ \cite{ThorneRMP}.
\bibitem{stationary}
By stationary here we mean that the perturbed spacetime is stationary
to first order, i.e. possesses a timelike killing field.
\bibitem{question}
{}From Eqs.\ (\ref{calE-curved}) -- (\ref{calB-down}) one would be
tempted to identify the
coefficients $a_{Jm}^{(2)}$ and
$c_{Jm}^{(2)}$ as the mass and current multipole moments of the star,
by analogy with the standard definitions in the asymptotically flat
case \cite{ThorneRMP}.  However, in our context the coefficient of
$r^{-J-3} T_{ij}^{2\,J+2,Jm}$ in ${\cal E}_{ij}^{(2)}$ is the sum of
$a_{Jm}^{(2)}$ and a contribution from the function ${}^{(\downarrow)}
{\cal E}_{ij}^{Jm}$.  Thus it is not clear that such an identification
would be valid.  The additional contribution scales like $(M/R)^{2
J+1}$ so is of a high post-Newtonian order.
\bibitem{notegauge}
The leading order pieces of ${}^{(\uparrow)} {\cal E}_{ij}^{Jm}$ and
${}^{(\downarrow)} {\cal E}_{ij}^{Jm}$ in an expansion in $M / r$ are
gauge invariant if one
demands that the gauge transformations have a regular limit as $M \to
0$, but not if one allows gauge transformations $\xi^{(2)\,a} \propto
1/M$.
\bibitem{noteT}
The explicit relation between our parameterization of the $J\ge 2$
piece of the solution
space and the function $\Psi_4$ can be obtained by combining
Eqs. (294) of Chapter 1 and (281) of Chapter 2 of Ref.\
\cite{Chandra}, and Eqs.\ (2.38e), (2.38f), and (2.33a) -- (2.33f) of
Ref.\ \cite{ThorneRMP}.  The result is
$$
R_{Jm}(r) = -{1 \over 4} \sqrt{ {(J+1) (J+2) \over (2 J -1 ) (2 J + 1)
}} \left( b_{Jm}^{(2)} + i d_{Jm}^{(2)} \right)
$$
$$
\times \ r^{J-2} \left[ 1 + O(M/r)\right]
$$
$$
-{1 \over 4} \sqrt{ {J (J-1) \over (2 J +1 ) (2 J + 3)
}} \left( a_{Jm}^{(2)} + i c_{Jm}^{(2)} \right) {1 \over r^{J+3}} \left[ 1 +
O(M/r)\right].
$$
\bibitem{note9}
This can be seen as follows.  The background metric
plus the spherically symmetric part of the perturbed metric, and the
fluid variables plus the spherically symmetric part of their
perturbations, form a static, spherically symmetric solution to the
Einstein--perfect fluid equations.  However, it is well known that
such solutions form a one
parameter family parameterized by total mass $M$ or total baryon
number $N$.  Therefore the spherically symmetric part of the perturbation
must consist of a displacement to a nearby static spherically
symmetric state, characterized by some non-zero shifts $\Delta M$ and
$\Delta N$.
\bibitem{proof1}
This can be shown, for example, by expanding the perturbation
variables in power series' about the center of the neutron star to
quadratic order, and deducing that there is no consistent choice of
coefficients for those power series when $c_{1m}^{(2)}(t) \ne 0$.
\bibitem{noteJ0}
In the analysis of Ref.\ \cite{Mino}, Mino et.\ al.\ set the $J=0$ part
of the $t \le 3$ perturbations to zero, arguing that ``it is natural
to suppose that the change of mass due to radiation reaction is
negligible''.  Our analysis here justifies this assumption by showing
that there is no $J=0$ part to the $t \le 3$ perturbations in the case
of a static spherical star; presumably the same is true for a black hole.
\bibitem{tailmetric}
The quantity that determines the $J=2$ piece of ${\hat h}_{ab}^{(3)}$
(aside from the piece that is linear in the first time derivative of
${\hat h}_{ab}^{(2)}$) is presumably the curvature of the ``tail
metric'' of Refs.\ \cite{Mino,QuinnWald} evaluated on the geodesic.
\bibitem{notePNGW}
The external scheme described here must be modified at
post-5/2-Newtonian order because of gravitational radiation.  The
standard way of performing this modification is to introduce a
post-Minkowskian expansion in a radiation zone and performing a
additional matched asymptotic expansion of the post-Newtonian fields
and the post-Minkowskian fields \cite{Burke,Blanchet}.  This
modification does not affect our arguments in Sec.\ \ref{sec:PN}.
\bibitem{pnc}
More details about the form of the post-Newtonian fields in this
coordinate-free description can be found in Ref.\ \cite{Tichy}.
\bibitem{commentlai}  Lai's analysis in Ref.~\cite{lai} actually
incorporates some first post-Newtonian effects, but not all of the
first post-Newtonian effects, and thus does not yield information
about the coefficient $F_1$.
\end{references}
\end{document}